# The effect of broadband soft X-rays in $SO_2$-containing ices: Implication on the photochemistry of ices towards young stellar objects

Pilling S.[1*], Bergantini A.[1]

[1]Universidade do Vale do Paraíba (UNIVAP), Laboratório de Astroquímica e Astrobiologia (LASA), São José dos Campos, SP, Brazil.

*e-mail: sergiopilling@pq.cnpq.br;

**ABSTRACT**

We investigate the effects produced mainly by broadband soft X-rays up to 2 keV (plus fast (~keV) photoelectrons and low-energy (eV) induced secondary electrons) in the ice mixtures containing $H_2O:CO_2:NH_3:SO_2$ (10:1:1:1) at two different temperatures (50 K and 90 K). The experiments are an attempt to simulate the photochemical processes induced by energetic photons in $SO_2$-containing ices present in cold environments in the ices surrounding young stellar objects (YSO) and in molecular clouds in the vicinity of star-forming regions, which are largely illuminated by soft X-rays. The measurements were performed using a high vacuum portable chamber from the Laboratório de Astroquímica e Astrobiologia (LASA/UNIVAP) coupled to the spherical grating monochromator (SGM) beamline at the Brazilian Synchrotron Light Source (LNLS) in Campinas, Brazil. *In-situ* analyses were performed by a Fourier transform infrared (FTIR) spectrometer. Sample processing revealed the formation of several organic molecules, including nitriles, acids, and other compounds such as $H_2O_2$, $H_3O^+$, $SO_3$, $CO$, and $OCN^-$. The dissociation cross section of parental species was in the order of $2-7 \times 10^{-18}$ cm$^2$. The ice temperature seems not to affect the stability for $SO_2$ in the presence of X-rays. Formation cross sections of produced new species were also determined. Molecular half-lives at ices towards YSOs due to the presence of incoming soft X-rays were estimated. The low obtained values, employing two different models of radiation field of YSOs (TW Hydra and typical T Tauri star), reinforce that soft X-rays are indeed a very efficient source of molecular dissociation in such environments.

## 1 INTRODUCTION

The UV and X-rays produced by stars may induce chemical changes in the surrounding matter (circumstellar and interstellar gas, solid-phase molecules (icy grains), planets, and other orbiting bodies) enhancing the chemical complexity in these environments (e.g. Tielens & Hollenbach 1985; Maloney, Hollenbach & Tielens 1996; Goicoechea et al. 2004; Henning & Semenov 2013; and references therein). Complex molecules have been detected by astronomical observations in such environments (e.g. Johnson & Quickenden 1997; Boogert et al. 2008; Gibb et al. 2000, 2004). In the case of new born stars, also called young stellar objects (YSOs), the material falling towards the stars makes the emission of X-rays even more intense than in stars in the main sequence (e.g. Siebenmorgen & Krügel, 2010; Imanish et al 2002). Such X-ray photons are capable of traversing large column densities of gas before being absorbed (Goicoechea et al. 2004; Casanova et al. (1995); Koyama et al. (1996); and Imanishi, Koyama & Tsuboi (2001). The X-ray-dominated regions (XDRs) in the interface between the ionized gas and the self-shielded neutral layers that surround the YSOs could influence the selective heating of the molecular gas and ice grains. The complexity of these regions possibly allows a combination of different scenarios and excitation mechanisms to coexist.

As discussed by Schleicher et al. (2010) and references therein, X-rays have much smaller cross section than UV photons and can thus penetrate larger columns. Specifically photons with energies around 1 keV penetrate a typical column of $2\times10^{22}$ cm$^{-2}$, a 10 keV photon penetrates $4\times10^{25}$ cm$^{-2}$, and high energies photons (hard X-rays) with energies around 100 keV can penetrates column density around $1\times10^{29}$ cm$^{-2}$. For comparison, a typical UV photon can penetrate roughly $\sim10^{21}$ cm$^{-2}$. Therefore, assuming that 2 keV X-rays (maximum photon energy employed in this work) can penetrates column densities up to $N_H < 10^{24}$ cm$^{-2}$ (Cecchi-Pestellini et al. 2009) and considering the relation between visual extinction and hydrogen column density, $A_V$ (mag) = 0.5 x $10^{-21}$ $N_H$ (cm$^{-2}$), or the relation between the reddening and hydrogen column density, E(B-V) (mag) = 0.15 x $10^{-21}$ $N_H$ (cm$^{-2}$) (Güver & Özel 2009), the maximum penetration depth of such X-rays corresponds to $A_v \sim$ 500 mag or E(B-V) $\sim$ 150 mag. In the scenario of YSOs, this large $A_v$ value indicates very dense regions, deep in the protostelar disk.

The photochemical complexity increases as the energy of the incoming photons increase, which lead neutral molecules to excited states, radicals, and ions at higher energies. In addition, in the case of X-rays, the produced high-energy ($\sim$ keV) photoelectrons and low-energy ($\sim$ eV) induced secondary electrons also represent an extra source for molecular processing. The domain of each set of reactions depends on the absorption cross section of each molecule involved. For example, as discussed by Pilling et al. (2009), in the case of ices containing $CH_4$ and $N_2$, the photochemical regime below 9 eV is governed by neutral-neutral or exited-neutral processes; for energies between approximately 10 and 14 eV, the chemical pathway involves neutral-radical as well as radical-radical; and for photon energies higher than $\sim$ 15 eV, the reaction involving ionic species dominates the photochemistry, which also increases the reaction rates due to the decrease of the activation barrier of the reaction routes.

Laboratory works have been performed in an attempt to simulate these chemical processing in astrophysical ices (or/and gas) employing different ionizing sources and analytical techniques (e.g. Gerakines et al. 2001; Chen. et al 2013; Jimenez-Escobar et al. 2012, Ciaravella et al. 2010, Boechat-Roberty, Pilling & Santos 2005; Pilling et al. 2006, 2007a, 2007b, 2009, 2011a, 2011b; Bernstein et al. 2004; Boechat-Roberty et al. 2009; Andrade et al. 2009; Fantuzzi et al. 2011). Recently, radiative-transfer models containing laboratory data in the infrared of processed ices (bombarded by cosmic-ray analog) were employed with success, to simulate the energy spectral distribution (SED) as well as the images, in different wavelength, of the young stellar object Elias 29 (Rocha and Pilling, 2015). The results obtained by the authors corroborate the idea of chemical changes induced by interaction between ices (as well as gas) with incoming ionizing radiation in such astrophysical environments. However, considering that the molecular abundances of astrophysical ices can largely vary from object to object (e.g. Gibb et al 2004), and only the most abundant species are well quantified in space, laboratory simulations of such ices always present a tiny point of view of the entire cosmological picture. For a detailed review of astrophysical ices see Boogert, Gerakines & Whittet (2015).

In this paper, we experimentally investigate, for the first time, the effects of ionizing broadband photons from 6 to 2000 eV (mostly soft X-rays and their induced fast ($\sim$ keV) photoelectrons and slow ($\sim$ eV) secondary electrons) in an ice mixture made of $H_2O:CO_2:NH_3:SO_2$ at two different temperatures (50 K and 90 K), which roughly simulates the physicochemical composition of some astrophysical ices. The current experiment is an attempt to simulate the photochemical processes induced by energetic photons in the $SO_2$-containing ices that are largely illuminated by soft X-rays around young stellar objects (e.g. Koyama et al. 1996).

As discussed before, the current investigation also simulates the effects of high energy photoelectrons ($\sim$ keV) and low-energy secondary electrons ($\sim$ eV) induced by soft X-rays photons in such astrophysical ices. This cannot be investigated when only UV lamps are used as the ionizing source. The employed instrumentation in this work is first reported here and will be subsidize in future experimental works.

## 2 EXPERIMENTAL SETUP

In an attempt to simulate the photochemical process triggered mainly by soft X-rays photons (broad band) on the $SO_2$-containing ices present in astrophysical environments, we used the facilities of the Brazilian Synchrotron Light Laboratory (LNLS) located in Campinas, Brazil. The experiments were performed using a high-vacuum portable chamber from the Laboratório de Astroquímica e Astrobiologia (LASA/UNIVAP) coupled to the spherical grating monochromator (SGM) beamline, which was operated in the off-focus and white beam mode, producing a wide spectral range of ionizing photons (mainly from 6 eV up to 2000 eV). The beamline details can be found elsewhere (Castro et al. 1997, Rodrigues et al. 1998). A gas mixture containing roughly 76% $H_2O$, 8% $CO_2$, 8% $NH_3$, 8% $SO_2$ was produced in a mixture chamber coupled to the experimental chamber. Two different sample temperatures were investigated (50 K and 90 K).

To better identify and quantify the new species produced during the sample processing involving the minor parental species present in the simulated ices, the amount of water in the samples was reduced roughly by a faction of ten when compared with astrophysical ices. Such reduction does not change the physical chemical process among the reagents because the water remains as a matrix for the other species. The description of the experiments is the following: first, the gaseous mixture (roughly 10 mbar) was slowly deposited into a polished ZnSe substrate (a dish with an effective area of 28.3 $mm^2$), previously cooled to 14 K by a helium closed cycle cryostat (ARS Inc., model CS204AB-450), inside a portable and high vacuum chamber (named "Stark" chamber). The gas mixture was deposited though a capillary stainless tube, held a few millimeters from the target for around 20 min, at a background pressure in the main chamber of $2x10^{-7}$ mbar. After deposition, the sample holder was slowly heated at 2 K $min^{-1}$ ramp up to the irradiation temperatures. Details of the analogue samples simulated in this work are shown in Table 1.

The atmosphere inside the chamber was monitored by quadrupole mass analyzer (MKS Inc., model e-Vision 2). The pressure inside the vacuum chamber during the irradiation was below $3x10^{-8}$ mbar which allows a small fraction of residual gas (mainly $H_2O$ and $CO_2$) over the sample. Considering the abundance of 1% of such compounds in the air, a unit sticking probability (i.e. S = 1) for both species, the layering happening in both sides of the ZnSe crystal, and the Langmuir unit (1 torr s = $10^{-15}$ molecules $cm^{-2}$ = 1 L = one monolayer), the estimated maximum value of layering of residual gas for ~ 700 minutes (initial heating+ irradiation + data acquisition stages) on the side that contained the ice sample was about ~ $5x10^{-15}$ molecules $cm^{-2}$ = 5 L. This value is about three orders of magnitude lower than the initial ice thickness given in terms of monolayers (4-5 kL).

*In-situ* chemical analyses of the samples were performed by a portable Fourier transform infrared (FTIR) spectrometer (Agilent Inc., model Cary 630) coupled to the experimental chamber. The spectra were taken from 4000 $cm^{-1}$ to 600 $cm^{-1}$ with resolution of 2 $cm^{-1}$.

Figure 1a presents a diagram of the experimental setup. A picture of the experimental hall of the Brazilian synchrotron light source (LNLS) with the Stark chamber coupled to the SGM beam line (arrow) is seen in Fig. 1b. Figure 1c shows, in detail, the frozen sample inside the Stark chamber ready to be irradiated by synchrotron light. The infrared (IR) beam from the FTIR and the synchrotron beam intercepted the sample perpendicularly, as illustrated in Fig. 1c. The infrared transmission spectra were obtained by rotating the substrate/sample by 90 degrees after each radiation dose and at selected temperatures during the heating stages. Infrared spectra of non-irradiated (pristine) samples taken at the beginning and at the end of the experiments were compared.

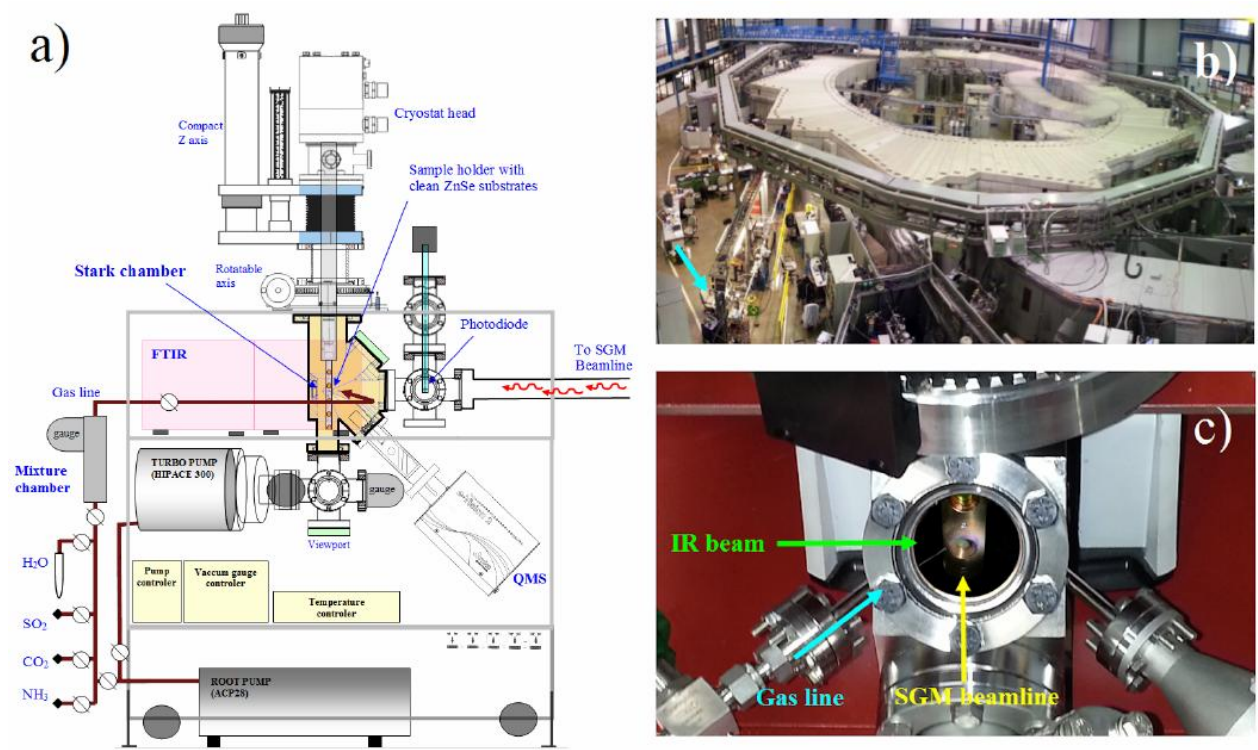

*Figure 1 - a) Diagram of the experimental setup (Stark chamber). b) Picture of the experimental hall of the Brazilian synchrotron source (LNLS) with the experimental chamber coupled at the SGM beam line (arrow). c) Picture showing the frozen sample inside the chamber and ready to be irradiated by synchrotron light. See details in the text.*

Table 1 lists key physicochemical parameters of the studied ices. The initial column density and thickness of the samples were determined by employing the methodology described in Pilling et al. (2011a, 2011b, and 2011c). The band strengths of $H_2O$, $CO_2$, $NH_3$, and $SO_2$ were taken from d'Hendecourt & Allamandola (1986), Gerakines et al. (1995), Kerkhof et al. (1999), and Garozzo et al. (2008), respectively. The initial sample thicknesses were between 1.5 and 1.8 μm. These values were estimated by measuring the area of the parent species in the IR spectra and adopting sample density of 1.0 g cm$^{-3}$.

**Table 1 - Samples characteristics.**

| Label | Sample composition | Temperature during the sample dosing (K) | Temperature during the irradiation (K) | Initial column density for $H_2O$, $CO_2$, $NH_3$, $SO_2$, respectively. ($10^{18}$ molecules cm$^{-2}$)[a] | Initial number of monolayes (L)[b] | Initial thickness (μm)[c] |
|---|---|---|---|---|---|---|
| E50K | $H_2O:CO_2:NH_3:SO_2$ (10:1.4:1.4:1.9) | 13K | 50K | 3.5, 0.5, 0.5, and 0.6 | 5200 | 1.8 |
| E90K | $H_2O:CO_2:NH_3:SO_2$ (10:2.4:2.3:1.7) | 13K | 90K | 2.5, 0.6, 0.6, and 0.7 | 4400 | 1.5 |

[a.] Employing the methodology described in Pilling et al. (2010a). Using the $H_2O$ band at 800 cm$^{-1}$ with band strength of 2.8 x$10^{-17}$ cm molecule$^{-1}$ (d'Hendecourt & Allamandola 1986); Using the $CO_2$ band at 2341 cm$^{-1}$ with band strength of 7.6 x$10^{-17}$ cm molecule$^{-1}$ (Gerakines et al. 1995); Using the $NH_3$ band at 1100 cm$^{-1}$ with band strength of 1.2 x$10^{-17}$ cm molecule$^{-1}$ (Kerkhof et al. 1999); Using the $SO_2$ band at 1329 cm$^{-1}$ with band strength of 3.7x$10^{-17}$ cm molecule$^{-1}$ (Garozzo et al. 2008). Molecular ratio was calculated before any irradiation and employing column densities at the sample temperatures of 50 K and 90 K, respectively.
[b] Considering the Langmuir unit (1 L = $10^{-6}$ Torr s ~ $10^{15}$ molecules cm$^{-2}$) and the summed molecular column densities in the ices in the beginning of each experiment.
[c] Employing methodology described in Pilling et al. (2011a, 2011b and 2011c) with density of 1.0 g cm$^{-3}$.

## 2.1 PHOTON FLUX

In this work, we simulate the concomitant effects of ionizing photons from vacuum ultraviolet (VUV) up to the soft X-ray range (dominant component). In terms of energy, it corresponds to the scope from 6 eV to 2 keV. Although photons with low energy (including near IR) reach the sample, only photons with energies higher than 6 eV promote significant changes such as photodesorption, photoexcitation, photoionization and photodissociation processes in the ice (e.g. Orlando & Kimmel 1997; Pilling et al. 2008), being even more intense for the soft X-ray component (> 100 eV). The broadband photon energy distribution (white beam mode) was obtained by placing the monochromator at zeroth order of reflection (the grating acts like an ordinary mirror and white light exits from the monochromator).

Figure 2 presents the photon flux of the SGM beamline as a function of photon energy (for the current of 200 mA and off-focus mode). The black squares are the measured values for the photon flux inside the experimental chamber in a monochromatic beam configuration. The red line is the model for the transmission of the SGM beam after passing through all optical elements using the SHADOWUII code. For comparison, four other photon fluxes are presented: i) non-attenuated photons from T-Tauri star at 1 AU - model 1 (adapted from the model of Siebenmorgen & Krügel, 2010); ii) estimated photon flux around the young star TW Hydra at 40 AU -model 2 (adapted from Fantuzzi et al. 2011); iii) solar flux at 1 AU (Gueymard, 2004) and at iv) 5.2 AU (adapted from Gueymard, 2004). The figure inset is the Chandra ACIS-I observation of T-Tauri star ROXs 21 at the photon energy from 0.5 to 7 keV (adapted from Imanishi et al. 2002), which indicates that soft X-rays are present in such objects. This figure illustrates that the SGM beam line is more related with the radiation distribution of YSOs than with stars in main sequence like the Sun, because the X-ray component produced by the falling material in the protostars is larger. In addition, the radiation distribution of young stellar objects in the soft X-ray range, as shown by the observation of T Tauri stars (e.g. Imanishi et al. 2002), has a maximum around 1 keV, which is very similar with the radiation profile of the employed synchrotron beamline. Although the unit of y-axis of the inset (observed spectrum) is different from the main figure, we superimposed the inset over the model of T Tauri type stars at 1 AU, respecting the x-axis, and assuming a correlation in y-axis at around 1 keV. In this figure we can also identify the different domains of ionizing photons in terms of energy (e.g. UV from ~3.3 to 6 eV; VUV form 6 to 100 eV; soft X-rays >100 eV).

The beamline entrance and exit slit were completely opened (L = 620 μm) during the experiments to allow the maximum intensity of the beamline. This implies energy steps of roughly 0.1 eV for the incoming photons in the sample. In an attempt to increase the beam spot at the sample, the experimental chamber was placed about 1.5 m away from the beamline focus (off-focus mode). With this procedure, the measured beam spot (rectangular) at the sample was about 0.5-0.6 cm$^2$. In addition, because the beamline was operated in the off-focus mode and the beamline slits were at maximum aperture, the homogeneity of photon flux inside this rectangular spot may be affected.

The integrated photon flux in the sample, corresponding mainly to photons from 6 eV to 2 keV, was $1.4 \times 10^{14}$ photons cm$^{-2}$ s$^{-1}$, and the integrated energy flux was roughly $7.6 \times 10^4$ erg cm$^{-2}$ s$^{-1}$. For comparison purpose, the integrated photon flux at the sample in the VUV range (6 to 100 eV) was estimated as roughly $4 \times 10^{13}$ photons cm$^{-2}$ s$^{-1}$, and the integrated energy flux in this wavelength range was roughly $3 \times 10^3$ erg cm$^{-2}$ s$^{-1}$. In the soft X-ray range (~0.1 to 2 keV), this flux was about $1 \times 10^{14}$ photons cm$^{-2}$ s$^{-1}$, and the integrated energy flux in this wavelength range was roughly $7 \times 10^4$ erg cm$^{-2}$ s$^{-1}$. The determination of the photon flux at the sample was done by the following procedure: i) measurement of the photon flux at selected energies (with bandpass of ΔE = 0.1 eV) in the soft X-rays by a photosensitive diode (AXUV-100, IRD Inc.) coupled to the experimental chamber (black squares in Figure 2); ii) scaling the theoretical beamline transmission flux, obtained by employing the XOP/SHADOWVUI ray-tracing code software (see http://www.esrf.eu/computing/scientific/xop2.0), using the measured photon flux at these specific photon energies. Absorption around 290 eV was due to contamination with carbonaceous molecules in the last mirror of the beamline. In the current experiments, we assume an average photon energy of ~ 1 keV, which gives the penetration depth up to L ~ 2 microns, considering the soft X-ray mass absorption coefficients for water (liquid phase) of μ ~ $4.1 \times 10^3$ cm$^2$ g$^{-1}$ taken from NIST database (see http://physics.nist.gov/PhysRefData/XrayMassCoef/) and an average sample density of 1 g cm$^{-3}$. The average penetration depth L of the employed ionizing photons was derived from the relation L= 1/μρ where μ is the mass absorption coefficient (determined mainly by atomic absorption cross sections) and ρ is the density of the material (e.g. Gullikson & Henke, 1989).

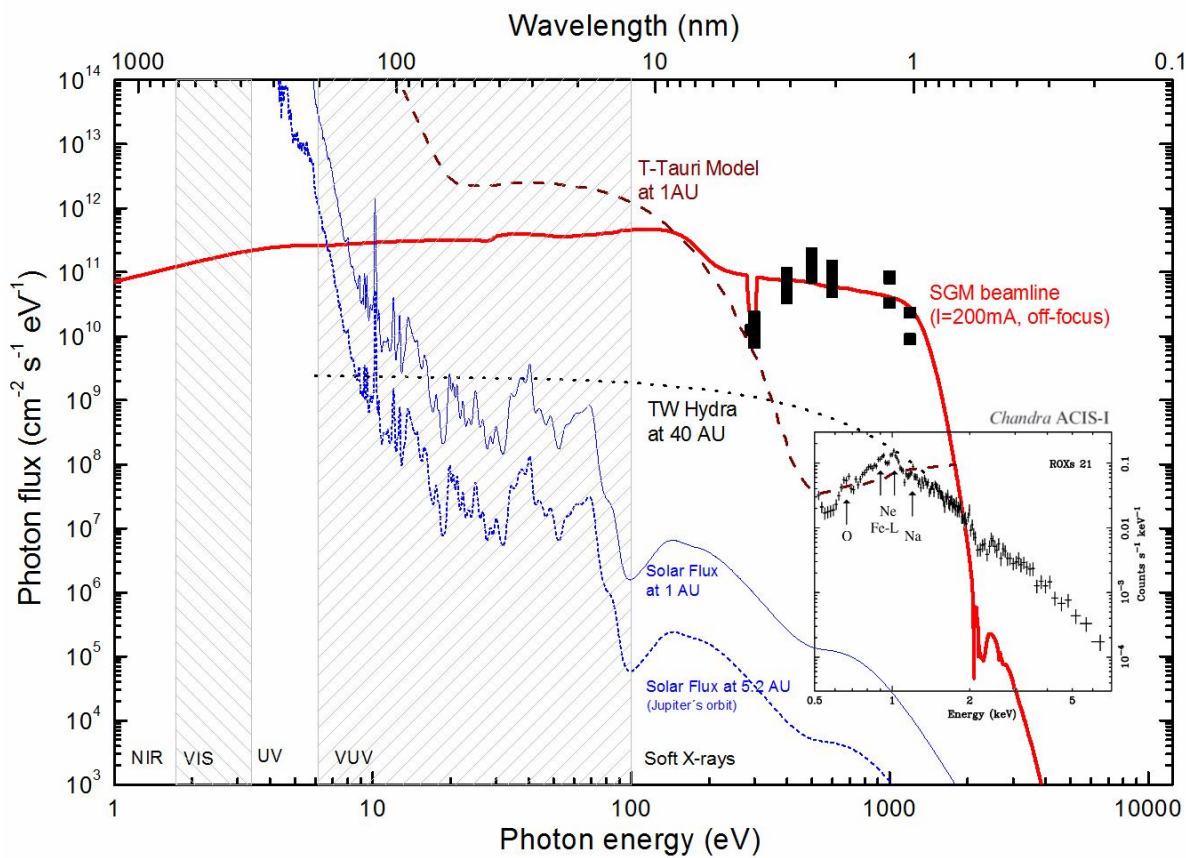

*Figure 2. a) Photon flux of the SGM beamline as a function of photon energy (for the current of 200 mA and off-focus mode). Black squares are the measured values for the photon flux inside the experimental chamber in a monochromatic beam configuration. The red line is the model for the transmission of the SGM beam after passing through all optical elements using the SHADOWUII code. For comparison, four other photon fluxes are presented: i) non-attenuated photons from T-Tauri star at 1 AU (adapted from the model of Siebenmorgen & Krügel, 2010); ii) Estimated photon flux around the young star TW Hydra at 40 (adapted from Fantuzzi et al 2011); iii) solar flux at 1AU and at iv) 5.2 AU (adapted from Gueymard, 2004). The figure inset is the Chandra ACIS-I observation of T-Tauri star ROXs 21 at the photon energy from 0.5 to 7 keV (adapted from Imanishi et al. 2002). See details in text.*

In this work, we focus on the astrophysical implications on the effects of the soft X-ray component of ionizing photons in the ices in the vicinity of young stellar objects. Here, only the radiation coming from the host star is considered; therefore, neither the effects of UV and X-rays coming from the interstellar environment, nor the ones originating from the interaction between cosmic rays and matter were considered. To take into account the fluctuation of the photon flux as a result of the natural energy loss (evolving currents from 250 mA to 130 mA) of the electrons that generate the synchrotron light and also their orbit corrections during the irradiations, in this paper, we adopt an average integrated photon flux from 6 eV to 2 keV, of about $1\times10^{14}$ photons cm$^{-2}$ s$^{-1}$, with an average energy flux of $6\times10^4$ erg cm$^{-2}$ s$^{-1}$ = $3.7\times10^{16}$ eV cm$^{-2}$ s$^{-1}$. This procedure introduces an estimated error of less than 20% in fluence determination of each collected IR spectrum.

Fewer X-ray irradiation experiments on astrophysical ices have been published, compared with other ionizing sources such as UV, fast ions, and electrons. The previous experiments performed by our group showed that X-rays promote efficient destruction rates (with cross section measurements) in astrophysical sample at both gas- and solid-phase (Boechat-Roberty, Pilling & Santos 2005; Pilling et al. 2006, 2007a, 2007b, 2009, 2011a, 2011b; Boechat-Roberty et al. 2009; Andrade et al. 2009; Fantuzzi et al.

2011). Jimenez-Escobar et al. (2012), measured the physicochemical changes induced by soft X-rays (0.3 keV) over pure $H_2S$ ice at 8 K. Although no cross sections were determined, the authors observed the appearance of $H_2S_2$ during the photolysis and also discussed the importance of X-rays on the processing of astrophysical ices outside and inside the solar system. An overview about the employment of soft X-rays in experimental astrochemistry is presented by Pilling and Andrade (2012).

**2.2 SOFT X-RAY-INDUCED PHOTOELECTRON AND SECONDARY ELECTRONS**

In general, as discussed by Gullikson and Henke 1989, X-ray absorption occurs via the photoelectric effect so that it results in the generation of a photoelectron with energy $E = h\nu - E_b$, where $h\nu$ is the soft X-ray photon energy (in this work $100 < h\nu$ (eV) $< 2000$ ) and $E_b$ is the binning energy of the electron. Electron binding energy is a generic term for the ionization energy that can be used for species with any charge state. For the studied compounds, the highest electron binding energy within the available photon energy range is found for O K-edge electrons (~ 540 eV), followed by N K-edge electrons (~400 eV), C K-edge electrons (~ 290 eV), and S L-edge electrons (~160 eV). The weakest electron binding energies come from the valence electrons of $_{NH3}$ (~10 eV), followed by valence electrons of $SO_2$ (~ 12.3 eV), $H_2O$ (12.6 eV), and $CO_2$ (~13.7 eV). The values above were taken from NIST gas phase ion energetics database (see http://webbook.nist.gov/chemistry/). The fastest photoelectrons produced by the most energetic incoming X-ray in the sample have kinetic energy around 1990 eV (e.g. photoelectron form single ionization of ammonia).

The effect of soft X-ray-induced secondary electron have been studied by several groups in the literature (e.g. Henke et al. 1979; Gullikson & Henke 1989, Akkermann et al. 1993; Hüfner 1995; Cazaux 2006; Pilling et al. 2009; Pilling & Andrade 2012*)* and is a rather complex process. As pointed out by Gullikson & Henke, 1989, a simplified description can be divided into four steps:

i) First, photoabsorption of X-ray and generation of energetic primary electrons (see above). Such fast primary electrons lose energy by creating electron hole (e-h) pairs at a distance that is generally short compared with the penetration depth L of the incoming photon (in this work we are considering L ~ 2 microns). Following Opal et al. 1972, the practical range of fast photoelectrons is related with its energy by $E^{1.9}$. The relation between the energies of primary photoelectron and the energy of subsequent secondary-electron for the parental species employed in this work (in gas phase only) is given elsewhere (Opal et al. 1972). From Monte Carlo simulations performed employing the CASINO code (see http://www.gel.usherbrooke.ca/casino/index.html), the range of 1 keV electrons is around 60 nm inside such astrophysical analog samples (Bergantini et al. 2014). For low-energy electrons (~ 5 eV), the mean free path inside such ices is around 1-2 nanometers.

ii) Excitation of secondary electrons by fast primary electrons (photoelectrons). Once a secondary electron is created, it will undergo a short random walk in the material while losing energy due to the creations of phonons (e.g. Gullikson & Henke 1989). Moreover, following Gullikson & Henke (1989), each soft X-ray may induce tens of secondary electrons inside matter, and this number increases as the temperature of the target increases. The mechanism involved in the generation of the secondary electrons results from the absorption of X-ray photons, photoelectric effect, followed by the creation of energetic photo and Auger electrons that generate a large number of low-energy secondary electron. For low atomic number, the fluxes of those secondary electrons inside matter can be considered as high as the incident X-rays. Although, for some Auger processes (e.g. satellite Auger) and shake up process, in which more than one electron can be ejected from the irradiated species, the flux can be even higher than the incident photon flux (see also Ramaker et al. 1988; Hüfner 1995; Almeida et al. 2014).

iii) Transport of the secondary electrons, including energy to the lattice. Following Henke et al. (1979), most of such secondary electrons have energies below 5 eV. When captured by a molecular target inside the bulk, such electrons can induce further molecular dissociating, for example via dissociative electron attachment mechanism ($AB + e^- \rightarrow AB^- \rightarrow A^- + B$). The mean free path of such secondary electrons is much reduced in comparison with photoelectrons, and its yield can vary depending on the sample density and temperature. Gullikson & Henke (1989), experimentally demonstrated, by employing soft X-rays (1487 eV) in Xe ices at very low temperatures, that the mean free path, escape depth, and secondary electron yields increase with the sample temperature.

iv) Escape of secondary electrons reaching the surface with sufficient energy to overcome any potential barrier at the surface or be thermalyzed/captured by any molecular species or center in the bulk. In astrophysical water-rich ices (insulators), a secondary electron will undergo many collisions before it either escapes through the surface or loses enough energy so that it is unable to escape or becomes trapped by interacting with a center in the bulk.

In the case of ices at the protostellar disk of YSO, the major component of incoming electrons has energies around keV and they are mostly produced from secondary processes during direct impact of cosmic rays matter (e.g. Bennett & Kaiser 2007). Following Kaiser 2002 and references therein, collisional cascade calculations demonstrate that each high energy (MeV) cosmic-ray particle can penetrate the entire astrophysical icy grain and induce cascades, generating up to hundreds of suprathermal particles such as these secondary electrons.

The employment of soft X-rays, to simulate the processing of astrophysical ices by radiation in comparison with similar bombardment experiments using fast electrons or cosmic ray analogs, helps us also to understand the effects of the secondary electrons inside matter. The production of secondary electrons is negligible when just UV or visible light is considered. Therefore, the produced energetic photoelectrons and low-energy secondary electrons discussed in this work will contribute to increase the chemical complexly (and its yields) of the irradiated samples and should be considered as an alternative source of electron-driven processes in such astrophysical ices.

Chen. et al (2013) and Ciaravella et al. (2010), in another set of experiments employing soft X-rays in astrophysical-related ices have also pointed out the importance of secondary electrons, whose energy distribution depends on the energy of incoming soft X-ray photons, in the production of new species in the ice. The authors irradiated frozen methanol with 0.3 and 0.55 keV photons and observed the production of $H_2CO$ due to chemical changing of ice triggered by incoming radiation.

## 3  RESULTS

As discussed before, the frozen samples were prepared at 13 K and were slowly heated (2 K min$^{-1}$) to the temperature of irradiation phase (50 K and 90K). We observe some chemical differentiation in sample for temperatures around 80 K. This physicochemical behavior is discussed in the Appendix and illustrated in Figure A1. A similar process of ice heating is commonly observed around YSOs, when we consider only non-ionizing photon (infrared, visible) delivered over grains as well as low-energy molecular adsorption (< 5 eV) and grain-grain collisions (e.g. turbulence) processes. However, as discussed by d´Hendecourt et al. (1982), Shen et al. (2004), and Ivlev et al. (2015), in the presence of high-energy ionizing radiation (e.g. UV, X-rays and cosmic rays) or very strong winds (e.g. supernovae), extra-heating mechanisms can also occur, allowing additional chemical differentiation in the grains. Both heating processes (in an extended period of time) allow total grain desorption/disruption delivering parent species as well as new products to gas phase (see details at Ivlev et al. 2015).

### 3.1 IRRADIATION PHASE AND REACTION ROUTES DURING ICE PROCESSING

The evolution of the infrared spectrum at 50K during the irradiation phase is shown in Figures 3a and 3b (expanded view). The bottom spectrum shows the unirradiated ice, and the uppermost spectrum was obtained at the highest photon fluence. Each spectrum has an offset for better visualization. Several new species, such as CO, $H_3O^+$, $SO_3$, $HSO_3$, $HSO_4^-$, and $SO_4^{-3}$, were identified during the photolysis. The $CO_3$ (2045 cm$^{-1}$) molecule was also formed, but it is not easily visible due to its low abundance. This species was also identified in similar irradiation and photolysis experiment of other $SO_2$-rich surface analogs, as discussed by Jacox & Milligan (1971). Table 3 lists the major absorption features observed in the infrared spectra of simulated astrophysical ices in this study (unirradiated sample, during the heating phase, during the irradiation phase, and in the residue at 300 K), with molecular attributions and comments.

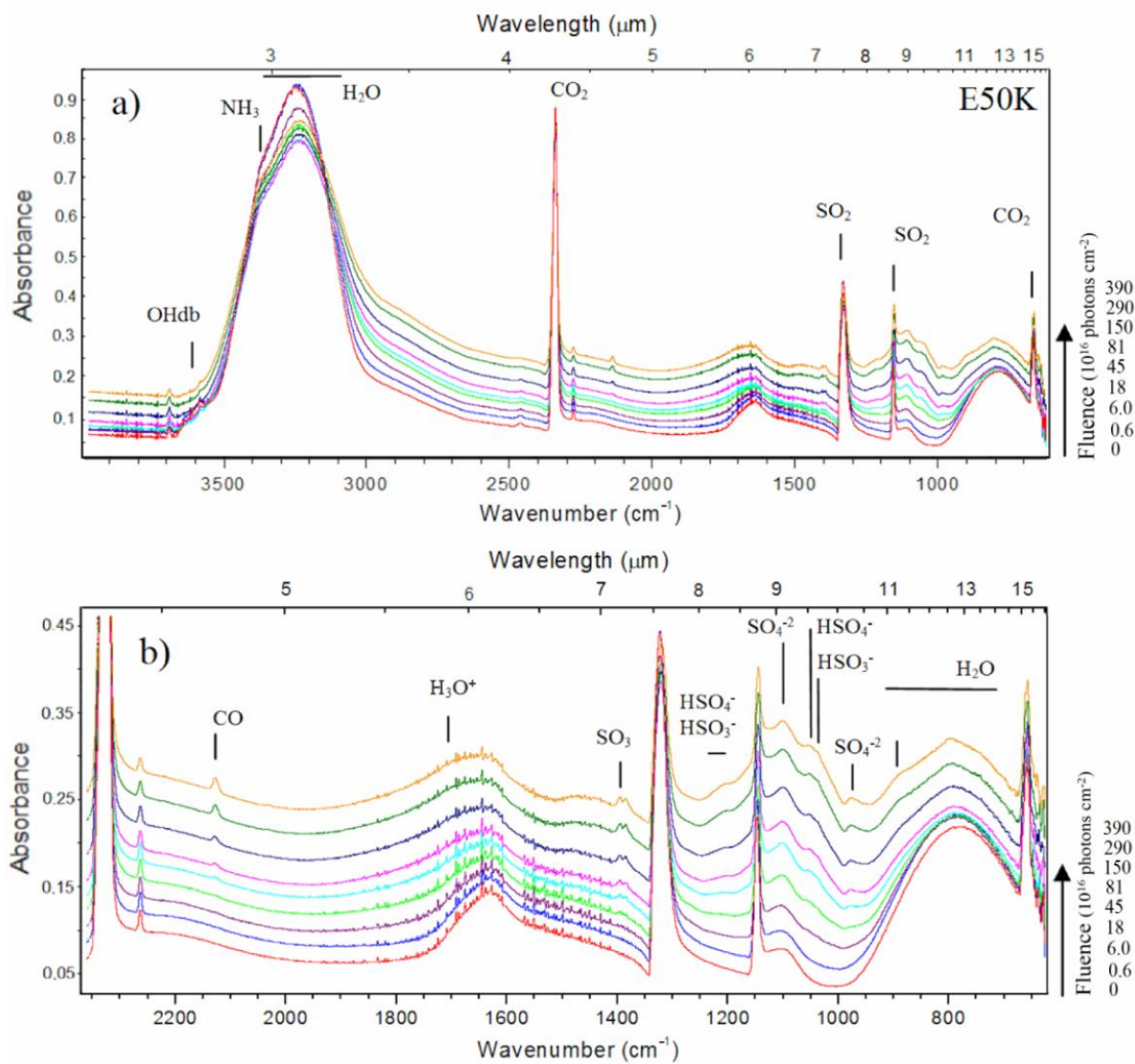

*Figure 3 - a) Infrared spectrum of the frozen sample at 50K during the isothermal irradiation, employing photons from 6 to 2000 eV (mainly soft X-rays), at different photon fluences. The bottom spectrum is the unirradiated ice, and the uppermost spectrum is the one obtained at the highest fluence. 5b) Expanded view from 2380 to 610 cm$^{-1}$. Each spectrum has an offset for better visualization.*

Figures 4a and 4b (expanded view) present the changes in the infrared spectrum of the sample irradiated at 90 K. In each figure, the bottom spectrum is the unirradiated ice and the uppermost spectrum is the one obtained at the highest photon fluence ($3.6 \times 10^8$ photons cm$^{-2}$). Each spectrum has an offset for better visualization. The spectrum at higher photon fluences shows the formation of cyanate ion, OCN$^-$ (2160 cm$^{-1}$), which was virtually not observed in the E50K experiment. Considering the band strength of $4.0 \times 10^{-17}$ cm molecule$^{-1}$ (d'Hendecourt, & Allamandola, 1986), at the fluence of $3.6 \times 10^8$ photons cm$^{-2}$, the column density for the produced OCN$^-$ was about $1.8 \times 10^{16}$ cm$^{-2}$. The clear observation of this species in the warm sample suggests that its reaction rate may mostly depend on the temperature. This temperature effect on the production of OCN- was also observed in the soft X-ray irradiation of another ice mixture $H_2O:NH_3:CO:CH_4$ (10:1:1:1) at two temperatures (20 K and 80 K) and will be reported in a future publication (Bergantini and Pilling, 2015, in preparation).

For the daughter species CO (2140 cm$^{-1}$), at this same fluence, the determined column density was about $1.2 \times 10^{16}$ cm$^{-2}$ considering the band strength of $1.1 \times 10^{-17}$ (Gerakines et al. 1995). The doublet line of SO$_3$ (1396 cm$^{-1}$; Schriver-Mazzuoli et al. 2003b), observed in the experiment E50K, was not clearly observed in this case. The decreasing of the OHdb band at around 3610 cm$^{-1}$ with the fluence is another feature observed in this figure. This effect is analogous to the compaction of amorphous ices during bombardment with fast ions (see also Palumbo 2006, Baragiola et al. 2008, and Pilling et al. 2010a). However, because the energy and momentum delivered by the impinging photons are small, we suggest that the decrease of the OHdb with the photon fluence is ruled by the dissociation (via X-rays and/or induced electrons) of dangling water inside micropores (see also Pilling et al. 2014; Bergantini et al. 2014; Gullikson and Henke 1989). Further experiments will help to clarify this issue.

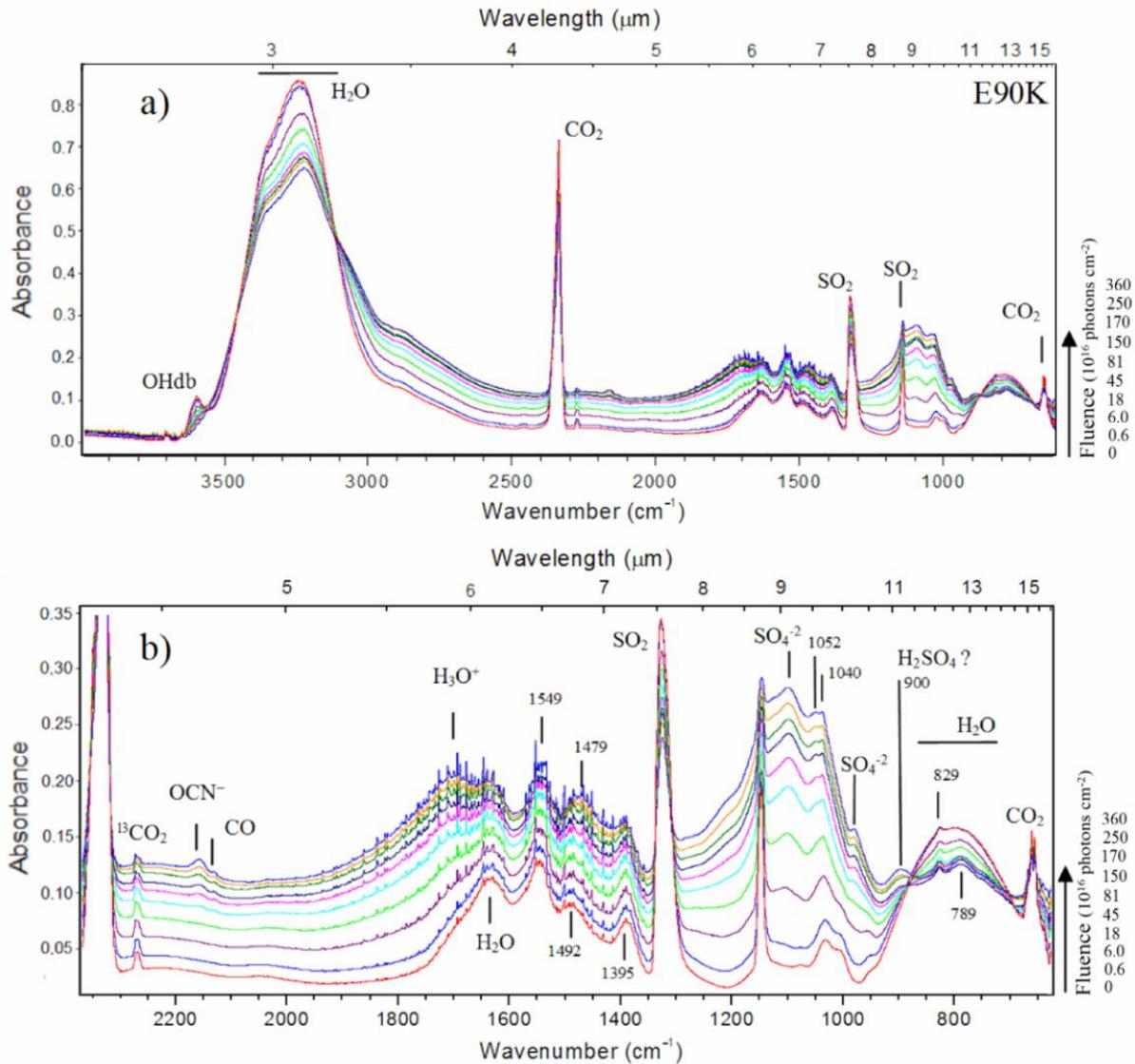

*Figure 4 - a) Evolution of infrared spectrum of the simulated samples at 90 K during the irradiation employing photons from 6 to 2000 eV (mostly soft X-rays). The bottom spectrum is the unirradiated ice, and the uppermost spectrum is the one obtained at the highest photon fluence. b) Expanded view from 2380 to 610 cm$^{-1}$. Each spectrum has an offset for better visualization. See details in the text.*

Figure 5a presents a comparison between selected infrared spectra of two studies ices. The upper spectra set shows the results for the experiments performed at 90 K and bottom spectra set shows the results of the experiments at 50 K. The labels in each spectrum indicate: a) deposited sample at 13 K, b) sample after slowly heating (2 K min$^{-1}$) up to the irradiation temperature (50 K or 90 K), and c) after irradiation up to the fluence of 3.9x10$^{18}$ photons cm$^{-2}$ (for E50K) and 3.6 x10$^{18}$ photons cm$^{-2}$ (for E90K).

Figure 5b shows an expanded view of the 2380 to 610 cm$^{-1}$ region in the infrared spectrum. Vertical lines indicate the bands that appear exclusively in the 90 K experiment. As discussed before, the production of OCN$^-$, observed by the 2160 cm$^{-1}$ infrared band, seems to be considerably enhanced in the sample irradiated at 90 K. The irradiation of the 90 K sample also caused a high destruction (or sublimation) of the trapped $CO_2$.

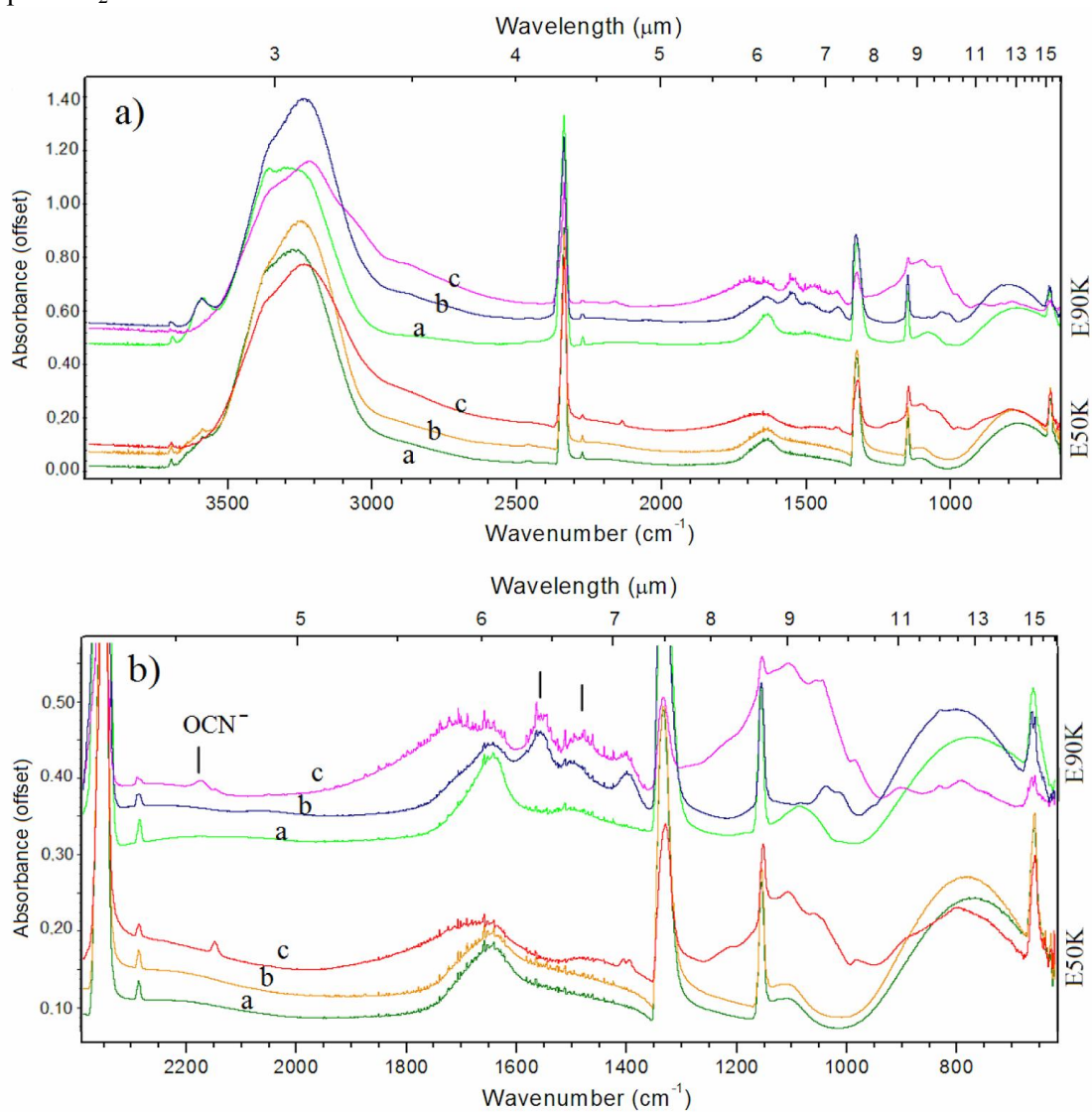

*Figure 5- a) Comparison between the experiments at 50 K (bottom spectra set) and 90 K (upper spectra set. The labels in each spectrum indicate: a) deposited sample at 13 K, b) sample after slowly heating (2 K min$^{-1}$) up to the temperature of the irradiation (50 K or 90 K), c) after irradiation up to the fluence of 3.9x10$^{18}$ photons cm$^{-2}$ (for 50 K) and 3.6x10$^{-18}$ photons cm$^{-2}$ (for 90 K). Figure b) shows an expanded view from 2380 to 610 cm$^{-1}$. Vertical lines indicate the bands that only appear in the experiment at 90K.*

**Table 3. Absorption features observed in the infrared spectra of simulated ices.**

| Wavenumber (cm$^{-1}$) | Wavelength (µm) | Molecule (vibration mode) | E50K | E90K | Notes and reference |
|---|---|---|---|---|---|
| 3610 (b) | 2.77 | OH db | unirradiated (w) | unirradiated | Pilling et al. 2010a |
| 3365 | 2.97 | NH$_3$ | unirradiated | unirradiated | Pilling et al. 2010a |
| 3226 | 3.10 | ? | residue (300K) | residue (300K) | |
| 3062 (b) | 3.27 | ? | residue (300K) | residue (300K) | |
| 2935 (w) | 3.41 | R-CH$_x$ ? | - | irradiated | |
| 2890 (w) | 3.46 | H$_2$O$_2$ | irradiated | irradiated | Pilling et al. 2010a, 2010b |
| 2868 | 3.49 | ? | - | residue (300K) | |
| 2857 | 3.50 | ? | residue (300K) | - | |
| 2466 | 4.06 | SO$_2$ (v1+ v3) | unirradiated | unirradiated | Khanna et al. 1988 |
| 2343 (s) | 4.27 | CO$_2$ | unirradiated | unirradiated | Pilling et al. 2010a |
| 2280 (w) | 4.39 | $^{13}$CO$_2$ | unirradiated | unirradiated | |
| 2166 | 4.62 | OCN- | - | irradiated residue (180K) | Pilling et al. 2010a |
| 2140 | 4.67 | CO | irradiated | irradiated | |
| 2045 (w) | 4.89 | CO$_3$ | irradiated | - | Jacox & Milligan 1971 |
| 1720 (b) | 5.81 | H$_3$O+ | irradiated | irradiated residue (180K and 300K) | Moore et al. 2007 |
| 1645 (b) | 6.08 | H$_2$O | unirradiated | unirradiated | |
| 1552 | 6.44 | ? RCOO$^-$ (carboxylate ion); zwiterionic amino acid ? | residue (170K) | unirradiated (90K) residue (180K) | Shimanouchi T., 1972 |
| 1493 | 6.70 | ? | - | unirradiated (90K) | |
| 1470 (b) | 6.80 | ? | irradiated | irradiated | |
| 1447 (b) | 6.91 | ? | - | residue (180K) | |
| 1422 (b) | 7.03 | ? | residue (300K) | residue (300K) | |
| 1396 (d) | 7.16 | SO$_3$ (v3) | irradiated | unirradiated (90K) | Schriver-Mazzuoli et al. 2003b |
| 1330 (s) | 7.52 | SO$_2$ (v3) | unirradiated | unirradiated | Khanna et al. 1988 |
| 1338 | 7.47 | ? | - | Residue (180K) | |
| 1294 | 7.73 | ? | Residue (300K) | -- | |
| 1255 | 7.97 | ? | Residue (170K) | - | |
| 1235 (b) | 8.10 | HSO$_4$- ; HSO3- | - | irradiated | Moore et al. 2007 |
| 1224 | 8.17 | H$_2$SO$_4$ | Residue (300K) | Residue (300K) | Moore et al. 2007 |
| 1215 | 8.23 | SO3 (poly) ? | irradiated | irradiated Residue (180K) | Moore 1984 |
| 1186 (b) | 8.43 | H$_2$SO$_3$ | - | Residue (300K) | Garozzo et al 2008 |
| 1169 | 8.55 | ? | Residue (300K) | - | |
| 1151 (s) | 8.69 | SO$_2$ (v1) | unirradiated | unirradiated | Khanna et al. 1988 |
| 1116 | 8.96 | ? | - | residue (300K) | |
| 1104 | 9.06 | SO$_4^{-2}$ (Sulfate); Hydrated H$_2$SO$_4$ ? | irradiated | irradiated | Moore et al. 2007; Strazzulla 2011 |
| 1103 | 9.07 | NH$_3$ | unirradiated | unirradiated | |
| 1101 | 9.08 | Hydrated H$_2$SO$_4$ ? | residue (300K) | - | Strazzulla 2011 |
| 1080 | 9.26 | SO$_4^{-2}$ | residue (170K) | residue (180K) | Zhang et al. 1993; Loeffler and Hudson 2013. |
| 1060 | 9.43 | SO$_3$ (v1) | residue (300K) | - | Moore 1984; Loeffler and Hudson 2013; Garozzo et al 2008. |
| 1052 | 9.51 | ? O$_3$; HSO$_4$-; | irradiated | irradiated | Moore et al. 2007 |

| | | H₂SO₃ | | | Garozzo et al 2008 |
|---|---|---|---|---|---|
| 1036 | 9.65 | ? O₃; HSO₃- (bisulfite) | irradiated residue (300K) | unirradiated (90K) Irradiated | Moore et al. 2007 |
| 1029 | 9.72 | ? O₃; H₂SO₃ | residue (300K) | residue (300K) | Schriver-Mazzuoli et al. 2003b; Garozzo et al 2008. |
| 1011 | 9.89 | ? | - | unirradiated (90K) | |
| 980 | 10.20 | SO₄²⁻ (Sulfate) | irradiated | irradiated | Moore et al. 2007 |
| 957 | 10.45 | H₂SO₄ ; S₂O₅⁻² | - | unirradiated (90K) | Moore et al. 2007; Loeffler and Hudson 2013. |
| 895(b) | 11.17 | ? | irradiated | irradiated | |
| 880(b) | 11.36 | ? H2SO4 | residue (170 and 300K) | residue (180 and 300K) | Moore et al. 2007 |
| 826(b) | 12.11 | SO₃ (poly) ? | - | unirradiated (90K) irradiated | Moore 1984 |
| 780 (b) | 12.82 | H₂O | unirradiated | unirradiated residue (180K) | Pilling et al 2010a; 2010b |
| 658 | 15.20 | CO₂ | unirradiated | unirradiated | Pilling et al 2010a; 2010b |

b-broad, n-narrow, w-weak, s-strong, d-doublet

The dissociation of the parental species by ionizing photons with energies within the 6 eV to 2000 eV range (mostly soft X-rays) may occur following several reaction routes, which may involve double dissociation or even triple dissociation, as well as neutral and ionic species (e.g. see discussion in Pilling et al. 2007c; Andrade et al. 2008). Some daughter species produced due to the interaction of ionizing photons with parental species in the ice samples are listed in the reactions below (asterisks indicate excited states; only neutral species are listed):

$H_2O + h\nu \rightarrow H_2O^* \rightarrow OH + H$ or $O + H_2$
$NH_3 + h\nu \rightarrow NH_3^* \rightarrow NH_2 + H$ or $NH + H_2$ or $N + H_2 + H$
$CO_2 + h\nu \rightarrow CO_2^* \rightarrow CO + O$ or $C + O_2$,
$SO_2 + h\nu \rightarrow SO_2^* \rightarrow SO + O$ or $S + O_2$,

Depending on the photon energy involved, some of the daughter species (neutral molecules, ions, or radicals) may diffuse inside the frozen sample and react with other species. As discussed by Loerting & Liedl (2000) and references therein, the formation of SO₃ from sulfur dioxide is highly enhanced in the presence of oxidant species such as hydroxyl (OH) and O₂, as illustrated by the reaction:

$SO_2 \xrightarrow{OH \text{ (or } O_2)} SO_3$

In the presence of water, the production of sulfurous acid (H₂SO₃) or sulfates (SO₄⁻²) is obtained by the reversible reaction:

$SO_3 + nH_2O \rightleftharpoons SO_3 \cdot nH_2O \rightleftharpoons H_2SO_4 \cdot (n-1)H_2O$.

However, as discussed by Strazzulla (2011), to obtain high yield of sulfurous acid in such reaction, the water concentration must not be greater than that of SO₂, otherwise the exceeding water would destroy the produced H₂SO₃, yielding again the SO₃ molecule. Another produced species observed is the bisulfite ion (or hydrogen sulfite) HSO₃⁻, which can be produced directly from SO₂ in the presence of frozen water molecules during a thermal heating process by the reaction:

$SO_2 + H_2O \rightleftharpoons HSO_3^- + H^+$.

During thermal heating or during irradiation of the sample, the sulfite ion $SO_3^{2-}$, which is the conjugate base of the bisulfite ion, can be produced by the reaction:
$HSO_3^- \rightleftharpoons SO_3^{2-} + H^+$.

Because all sulfites (including bissulfite) and sulfur dioxide, which contains sulfur in the same oxidation state (+4), are reducing agents, they react with the free hydrogen atoms (and protons) produced from the dissociation of water due to ionizing radiation. This mechanism may reduce the amount of these atoms in the sample, thus reducing the amount of produced hydrocarbons in the residue.

## 3.2 CROSS SECTION DETERMINATION

Figure 6a shows the evolution of the difference spectra (spectrum at a given fluence subtracted from the spectrum of the unirradiated sample) for the E50K experiment at several fluences. Peaks pointing upward (i.e. with positive area) indicate the production of new species. The downward pointing peaks (negative area) indicate the destroyed species (i.e. parental species).

The numerical evolution of the abundance of a given molecular species in the ice during the irradiation (parent or daughter) can be quantified by the equation

$$A - A_o = A_\infty \cdot (1 - \mathrm{EXP}(-\sigma_{d,f} \cdot F)) \qquad [\mathrm{cm}^{-1}] \qquad [1]$$

where $A$, $A_o$, and $A_\infty$ are the selected area of infrared band related with a specific vibrational mode of a given molecule at a given fluence ($A$), at the beginning of the experiment ($A_0$, unirradiated sample), and at the highest fluence ($A_\infty$, terminal fluence). In this expression, $\sigma_{d,f}$ represents the formation cross section ($\sigma_f$) for new species (daughter), or the destruction cross section ($\sigma_d$) for parental species, both in units of $\mathrm{cm}^2$, depending on each case. F indicates the photon fluence in units of $\mathrm{cm}^{-2}$. Moreover, for parental species, at the terminal fluence, Eq. 1 shows that the value of $A-A_o \rightarrow A_\infty \neq -A_o$ (i.e. $A(F_\infty) \neq 0$). This may be associated with three different issues: i) An incomplete soft-X ray illumination area in comparison with the probed sample by the IR beam. As discussed before, the outer border of the frozen sample may not have been fully illuminated by the incoming soft X-rays due to an eventual inhomogeneity of the beamline spot. ii) The incoming radiation only illuminated the upper layers of the sample with the bottommost layers constantly shielded. iii) The destroyed and produced species reached a photochemical equilibrium after a given fluence. Future investigation will help to clarify this issue.

Figure 6b shows the evolution of difference band areas (band area at a given fluence subtracted by the band area of the unirradiated sample) for selected new species (upper panel) and for parental species (bottom panel) as a function of photon fluence for the E50K experiment. The lines indicate the best fit for the experimental data, employing Eq. 1 for both produced and destroyed species. The determined cross sections are also shown in this figure. Figures 7a and 7b present similar results for the experiment E90K.

The determined values for the dissociation cross section of parental species and the formation cross sections of selected daughter species from the studied ices under the influence of ionizing photons between 6-2000 eV, are given in Tables 4 and 5, respectively. The destruction cross section of OHdb band in the E50K and E90K experiments are $5 \times 10^{-18}$ $\mathrm{cm}^2$ and $8 \times 10^{-18}$ $\mathrm{cm}^2$, respectively. An extended discussion about this will be given in the next section. Such values of cross section are in good agreement with previous measurements employing soft X-rays in other astrophysical-related samples such as acetone ice at 10 K (Almeida et al. 2014), $N_2$:$CH_4$ (19:1) ice at 15 K (Pilling et al. 2009), pyrimidine ice at 130 K (Mendoza et al 2013), solid-phase amino acids and nucleobases at 300 K (Pilling et al. 2011a), carboxylic acids (Boechat-Roberty et al. 2005; Pilling et al. 2006), methanol (Piling et al. 2007a), benzene (Boechat-Roberty 2009), and methyl formate (Fantuzzi et al. 2011).

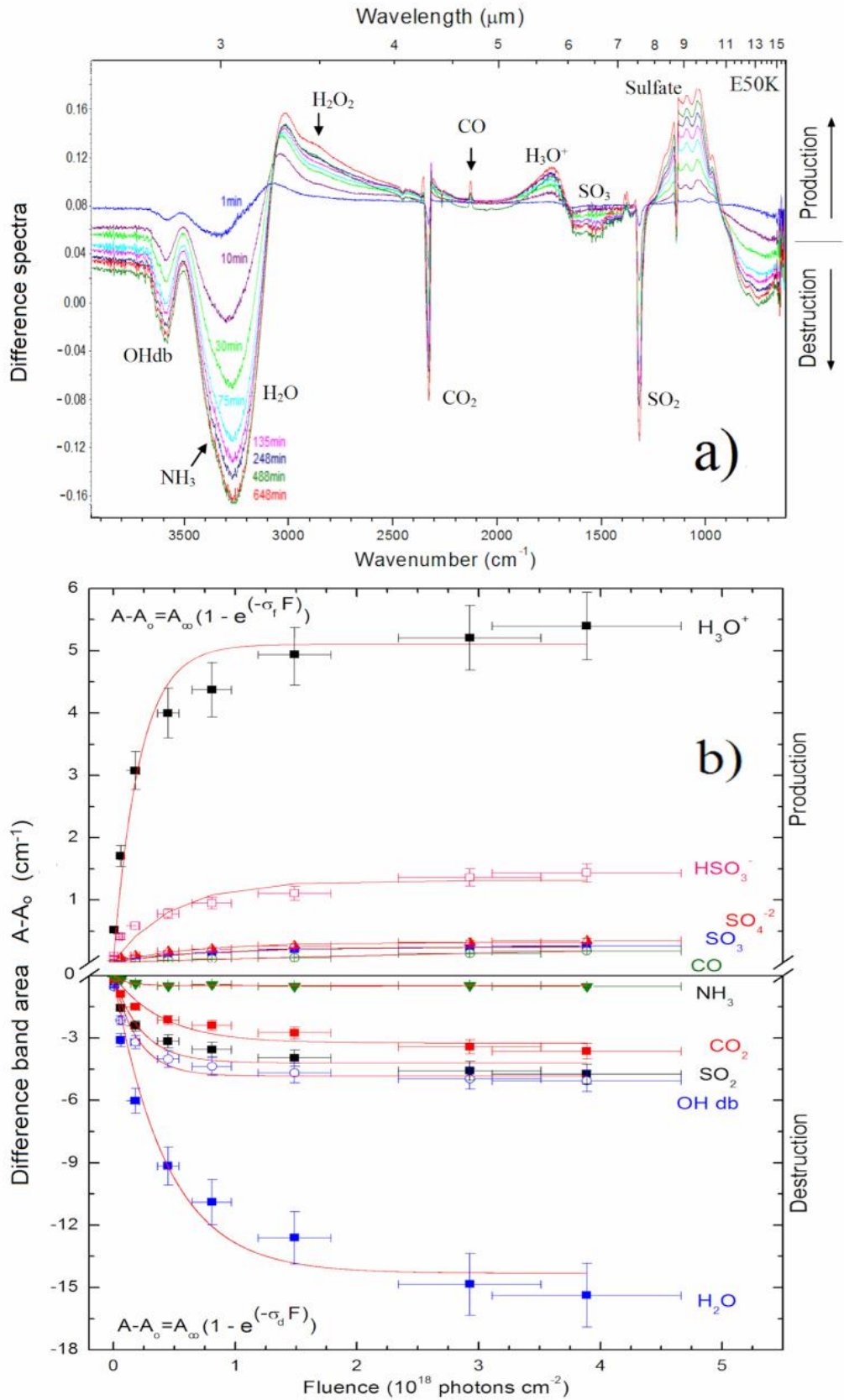

*Figure 6- a) Difference spectra for E50K experiment. Positive values for absorbance indicate the vibration modes of new products in the sample. b) Subtracted peak area for selected new species (upper panel) and parental species (bottom panel), as a function of photon fluence for the same experiment. Lines indicate the best fit of data employing Eq 2.*

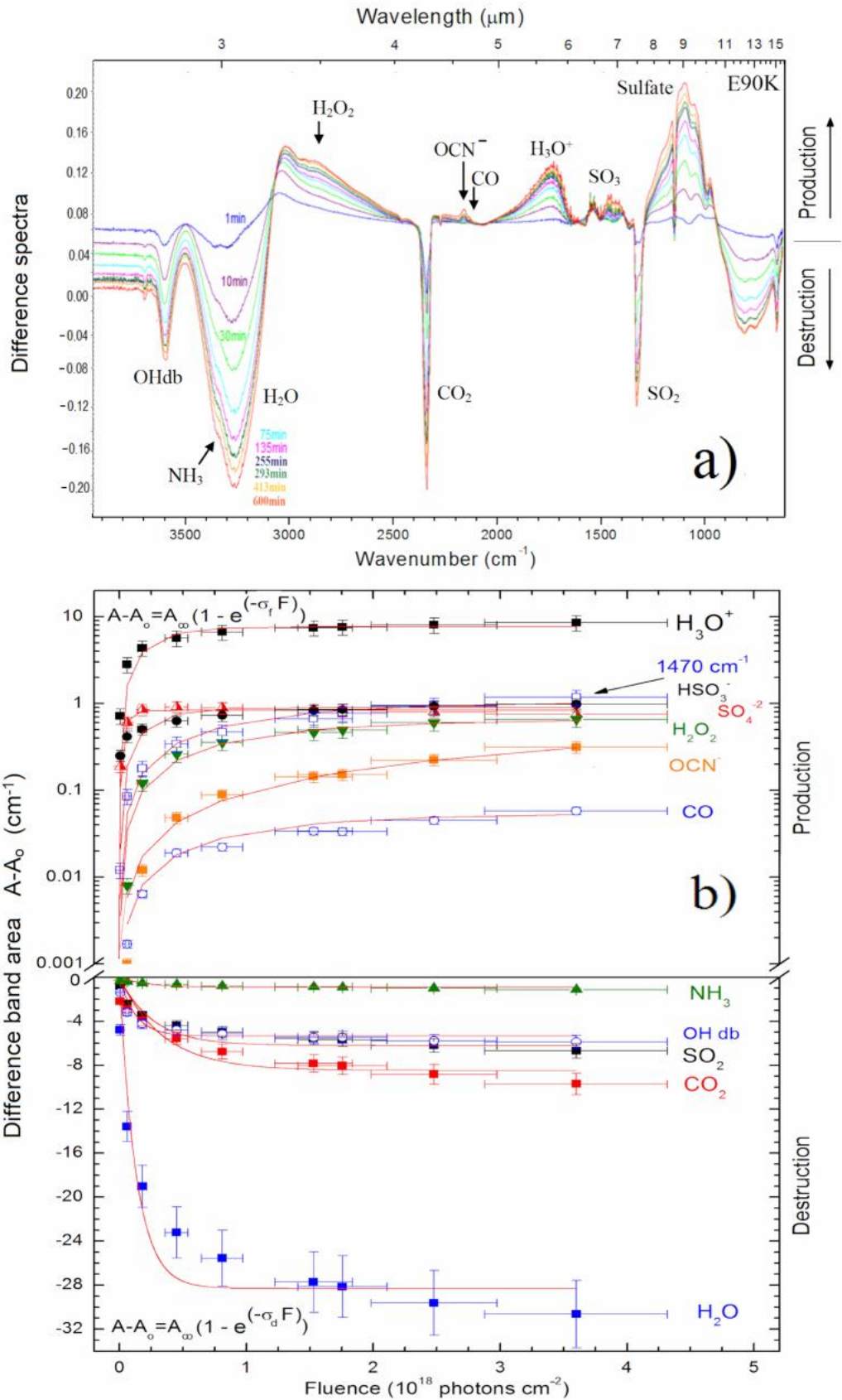

*Figure 7 - a) Difference spectra for E90K experiment. Positive values for absorbance indicate the vibration modes of new products in the sample. b) Subtracted peak area for selected new species (upper panel) and parental species.*

From the dissociation/formation cross sections, it is also possible to calculate the dissociation/formation rate by the equation:

$$k = \sigma \times \phi \quad [s^{-1}] \quad [2]$$

where $\phi$ indicates the photon flux in units of photons cm$^{-2}$ s$^{-1}$. The dissociation and formation rates for the studied species due to irradiation of photons with energies from 6 to 2000 eV in the laboratory (considering $\phi_{lab}$ = 1 x10$^{14}$ photons cm$^{-2}$ s$^{-1}$) are also listed in Table 4 and 5. The half-life of parental species during the experiments was determined by the expression

$$t_{1/2} \text{ (lab)} = \ln(2)/k \quad [s] \quad [3]$$

where k is the photodissociation rate in units of s$^{-1}$. These values are also listed in the Table 4.

**Table 4 - Dissociation cross sections and dissociation rate for parental species in studied ices irradiated by photons with energies from 6 to 2000 eV (mostly soft X-rays). Half-life obtained in the lab is also given. The uncertainty was estimated to be around 20%**

|  | E50K | | | E90K | | |
| --- | --- | --- | --- | --- | --- | --- |
|  | $\sigma_d$ (cm$^2$) | $k_{lab}$ (s$^{-1}$) | $t_{1/2}$ (lab) [a] [$10^3$ s] | $\sigma_d$ (cm$^2$) | $k_{lab}$ (s$^{-1}$) | $t_{1/2}$ (lab) [a] [$10^3$ s] |
| $H_2O$ [b] | 3.0x10$^{-18}$ | 3x10$^{-4}$ | 2.3 | 7.0x10$^{-18}$ | 7.0x10$^{-4}$ | 0.9 |
| $CO_2$ | 2.2x10$^{-18}$ | 2.2 x10$^{-4}$ | 3.1 | 3.0x10$^{-18}$ | 3.0 x10$^{-4}$ | 2.3 |
| $NH_3$ | ~6x10$^{-18}$ | ~6 x10$^{-4}$ | ~1 | ~2x10$^{-18}$ | ~2x10$^{-4}$ | ~3 |
| $SO_2$ | 4.3x10$^{-18}$ | 4.3 x10$^{-4}$ | 1.6 | 4.0x10$^{-18}$ | 4.0x10$^{-4}$ | 1.7 |

[a] Considering the half-life $t_{1/2}$ = ln(2)/k, where k is the photodissociation rate in units of s$^{-1}$ (see Table 4).
[b] For OHdb band the destruction cross sections are 5 x10$^{-18}$ cm$^2$ and 8 x10$^{-18}$ cm$^2$ for E50K and E90K experiments, respectively.

**Table 5 - Formation cross sections and formation rate for selected daughter species in the studied ices irradiated by photons with energies from 6 to 2000 eV. The uncertainty was estimated to be around 20%.**

| Molecule | E50K | | E90K | |
| --- | --- | --- | --- | --- |
|  | $\sigma_f$ (cm$^2$) | $k_{lab}$ (s$^{-1}$) | $\sigma_f$ (cm$^2$) | $k_{lab}$ (s$^{-1}$) |
| $H_3O^+$ | 4.6E-18 | 4.6x10$^{-4}$ | 4x10$^{-18}$ | 4x10$^{-4}$ |
| $H_2O_2$ | - | - | 9x10$^{-19}$ | 9x10$^{-5}$ |
| $HSO_3^-$ | 2.13x10$^{-18}$ | 2.1x10$^{-4}$ | 4x10$^{-18}$ | 4x10$^{-4}$ |
| $SO_3$ | 1.33x10$^{-18}$ | 1.3x10$^{-4}$ | - | - |
| $SO_4^{-2}$ | 1.63x10$^{-18}$ | 1.6x10$^{-4}$ | 2x10$^{-17}$ | 2x10$^{-3}$ |
| CO | 1.63x10$^{-19}$ | 1.6 x10$^{-5}$ | 9x10$^{-19}$ | 9x10$^{-5}$ |
| $OCN^-$ | - | - | 7x10$^{-20}$ | 7x10$^{-6}$ |
| 1470 cm$^{-1}$ | - | - | 9x10$^{-19}$ | 9x10$^{-5}$ |

From previous experiments employing different ices and ionization sources, we observe that the energy and the type of ionization source, be it photons, electrons, or ions, play the most important role in the dissociation of molecules (e.g. Pilling et al. 2010a, 2011a, and 2014). However, depending on the presence of given reactive parental species in the sample, such as the $SO_2$, the dissociation mechanisms of adjacent species in the ice may be affected due to the opening of different reaction pathways involving reactive ionic species and radicals produced by the incoming radiation. Further experiments with different molecular concentrations of parental species may help to elucidate this issue.

## 3.3 ENERGY FLUENCE TO REACH HALF-LIFE

The energy fluence to reach the half-life $D_{1/2}$ (sometimes called radiation dose to reach the half-life or half-life dose), obtained at the fluence in which the molecular abundance of a given parent species reaches half of the its initial value in the lab, is derived from the equation

$$D_{1/2} = t_{1/2} \times E \qquad [\text{eV cm}^{-2}] \qquad [4]$$

where and $t_{1/2}$ is the half-life in units of seconds and $E$ is the integrated energy flux in units of eV cm$^{-2}$ s$^{-1}$ of the incoming radiation (taken from Pilling et al. 2014).

The radiation dose to reach the half-life in the laboratory for the studied species, is listed in Table 6. Considering the average penetration depth of the employed ionizing photons L ~ 2000 nm and the sample density of 1 g cm$^{-3}$, the estimated amount of "illuminated" molecules inside the sample is around $1.2 \times 10^{17}$ molecules, as a first approximation, in a cylinder-shape volume with base area of 1 cm$^2$ (samples area) and height of 2000 nm (see equations in Pilling et al. 2012). From the amount of molecules exposed to radiation, we can also derive the energy fluence to reach the half-life in units of eV/molecule. This quantity is also listed in Table 6.

**Table 6 - Radiation dose to reach the half-life ($D_{1/2}$) in the laboratory during irradiation with photons from 6 to 2000 eV.**

|  | E50K | | E90K | |
| --- | --- | --- | --- | --- |
|  | $D_{1/2}$ (lab)[a] [$10^{20}$ eV/cm$^2$] | $D_{1/2}$ (lab)[b] [$10^4$ eV/molecule] | $D_{1/2}$ (lab)[a] [$10^{20}$ eV/cm$^2$] | $D_{1/2}$ (lab)[b] [$10^4$ eV/molecule] |
| $H_2O$ | 25 | 2.1 | 4 | 10.3 |
| $CO_2$ | ~ 85 | ~7.5 | 10 | 1 |
| $NH_3$ | 4 | 0.3 | 90 | 7.8 |
| $SO_2$ | 10 | 0.8 | 7 | 0.6 |

[a] Obtained directly from experimental data at the energy fluence in which the numeric abundance of a specific molecular species reaches roughly the half-value.
[b] Considering the average penetration depth of ionizing photons from 6 to 2000 eV of L ~ 2 microns and the sample density of 1 g cm$^{-3}$, the estimated amount of molecules illuminated in the sample inside a cylinder-shape volume with base area of 1 cm$^2$ and height of 2000 nm is calculated to be around $1.2 \times 10^{17}$ molecules (see equations at Pilling et al 2012).

## 4 – DISCUSSION AND ASTROPHYSICAL IMPLICATIONS

Generally speaking, astrophysical ices are made by the freeze-out of gas phase elements or small molecules (mostly $H_2O$, $NH_3$, $CO_2$, CO, $CH_3OH$, $SO_2$, and more) onto cold grains (made initially by refractory species such as oxides, silicates and amorphous carbon - grain seed) in dense and cold e interstellar and circumstellar environments. As discussed before, the abundances of such species in the ices largely vary depending on the environments. However, there is a consensus in the literature about the average ratio of the most abundant species, $H_2O$ and $CO_2$. For example, following Gerakines et al. (1999), the $CO_2/H_2O$ ratio seems to be around 19% in the vicinity of massive YSOs and 30% or higher toward low mass YSOs. A detailed review of astrophysical ices can be seen at Boogert, Gerakines & Whittet (2015).

The presence of $SO_2$ in star-forming regions, such as Orion, Sgr B2, and in the vicinity of YSOs, have been suggested by several authors (Snyder et al. 1975; Turner 1995; Boogert, et al. 1996; Boogert, et al 1997; Schilke et al. 2001; Keane et al. 2001; Boogert et al. 2002; Gibb et al 2004; Comito et al. 2005; Caux et al. 2011;). As discussed by Lattanzi et al. (2011), molecules with sulfur account for about 10% of the species identified in the interstellar gas and in circumstellar envelopes, and the $SO_2$ is one of the most prominent of these species. Zasowski et al. (2009) tried to quantify the presence of $SO_2$ in the ices toward YSOs using infrared observations from Spitzer. Following the authors, the average composition of the ices toward Class I/II YSOs was dominated by water, with 12% $CO_2$ (abundance relative to $H_2O$), 14% $NH_3$, 0.5% $SO_2$ and other compounds. However, as discussed by Boogert et al. 1996, in the case of W33A, a high mass YSO, the relative abundance of $SO_2$, in respect to water, was determined to be about 1.6%. Depending on the position of such ices in the circumstellar environment of YSOs, for example, if embedded or not in the prostostellar disk, their temperatures may be as low as ~20 K for the high embedded grains (Av > 8 mag), or as high as 50-160 K (Av ~ 3 mag) for the intermediated layers (midplane) of the disk (e.g. Henning & Semenov 2013; Öberg et al. 2011). Therefore, when applied to YSOs, the employed temperatures in this work (50 K and 90 K) roughly simulated the ices at different regions in the midplane of the disk fully illuminated by X-rays.

As discussed before, due to the uncertainties in the real composition of astrophysical ices as well as due to its diversity, depending on the studied object, each laboratory simulation only gives a small glimpse into the big cosmological picture. Therefore, the current simulated ices (mixed ice $H_2O:CO_2:NH_3:SO_2$ with ratio around 10:1:1:1 and temperatures of 50 K and 90 K) and the radiation field employed (broadband soft X-rays) help us to understand the physicochemical processes over $SO_2$-containing ices in the presence of ionizing radiation field in these environments (which can also vary) rather than simulating specific YSOs.

Table 7 presents, for the sake of comparison, the integrated photon flux and energy flux measured in the lab at different photon energy ranges, as well as some values adopted for two ionization models of YSOs: model 1 (adapted from Siebenmorgen & Krügel 2010) and model 2 (from Fantuzzi et al. 2011) considered in this work. The solar flux 1 AU and at 5.2 AU (Jupiter´s orbit) are also given (adapted from Gueymard, 2004).

**Table 7 - Integrated photon flux and energy flux at different photon energy range considered in this work.**

Integrated photon flux ($cm^{-2} s^{-1}$)

| Photon energy range (eV) | Lab [a] | Model 1 Typical YSO at 1AU [b] | Model 2 TW Hydra at 40 AU [c] | Sun at 1AU [d] | Sun at 5.2 AU [d] |
|---|---|---|---|---|---|
| 6-100 | $3.6 \times 10^{13}$ | $\sim 5.7 \times 10^{15}$ | $2.0 \times 10^{11}$ | $\sim 1.9 \times 10^{13}$ | $\sim 4.6 \times 10^{11}$ |
| 100-2000 | $9.7 \times 10^{13}$ | $7.1 \times 10^{13}$ | $7.3 \times 10^{11}$ | $8 \times 10^{8}$ | $3 \times 10^{7}$ |
| 6-2000 | $1.3 \times 10^{14}$ | $5.8 \times 10^{15}$ | $9.3 \times 10^{11}$ | $1.9 \times 10^{13}$ | $4.6 \times 10^{11}$ |

Integrated energy flux ($eV\ cm^{-2} s^{-1}$)

| Photon energy range (eV) | Lab [a] | Model Typical YSO at 1AU [c] | Model 2 TW Hydra at 40 AU [b] | Sun at 1AU [d] | Sun at 5.2 AU [d] |
|---|---|---|---|---|---|
| 6-100 | $0.1 \times 10^{16}$ | $5.4 \times 10^{16}$ | $\sim 9.9 \times 10^{12}$ | $\sim 6.2 \times 10^{13}$ | $\sim 4.3 \times 10^{12}$ |
| 100-2000 | $3.6 \times 10^{16}$ | $8 \times 10^{15}$ | $3.4 \times 10^{14}$ | $2.5 \times 10^{11}$ | $1.5 \times 10^{10}$ |
| 6-2000 | $3.7 \times 10^{16}$ | $6.2 \times 10^{16}$ | $3.5 \times 10^{14}$ | $6.3 \times 10^{13}$ | $4.4 \times 10^{12}$ |

[a.] *SGM beamline at the Brazilian Synchrotron Light source (white beam mode). At sample position.*
[b.] *Non-attenuated photons from T-Tauri star at 1 AU (Adapted from the model of Siebenmorgen & Krügel, 2010).*
[c.] *Adapted from Fantuzzi et al 2011.*
[d.] *Solar flux at 1AU and a 5.2 AU (Adapted from Gueymard, 2004).*

An estimation for the photons flux (6 to 2000 eV) inside YSOs as a function of visual extinction (Av) employing both ionization models are presented in Figure 8a. For model 1, we considered the non-attenuated flux $\phi_x$ (Av=0) = $5.8 \times 10^{15}$ photons $cm^{-2} s^{-1}$ for typical T Tauri star at 1 AU (adapted form Siebenmorgen & Krügel, 2010). For model 2, we adopt $\phi_x$ (Av=0) = $1.5 \times 10^{15}$ photons $cm^{-2} s^{-1}$ for TW Hydra at 1 AU (adapted from Fantuzzi et al. 2011). As a first hypothesis, we assume that all the photons in this energy range are roughly equally attenuated by the gas/dust, and the considered relation between photon flux and visual extinction is given by the expression

$$\phi_x(Av) \sim \phi_x(Av=0)\, e^{-0.9\, Av} \qquad [\text{photons cm}^{-2}\ \text{s}^{-1}] \qquad [5]$$

where $\phi_x(Av=0)$ represents the non-attenuated photon flux in units of photons cm$^{-2}$ s$^{-1}$ and Av is the visual extinction (considered dimensionless in this equation) (adapted from Mendoza et al. 2013). In the calculations above, the central stars are considered the only source of X-rays, and we adopt the r$^{-2}$ dependence with the distance as dilution factor for 10 and 100 AU.

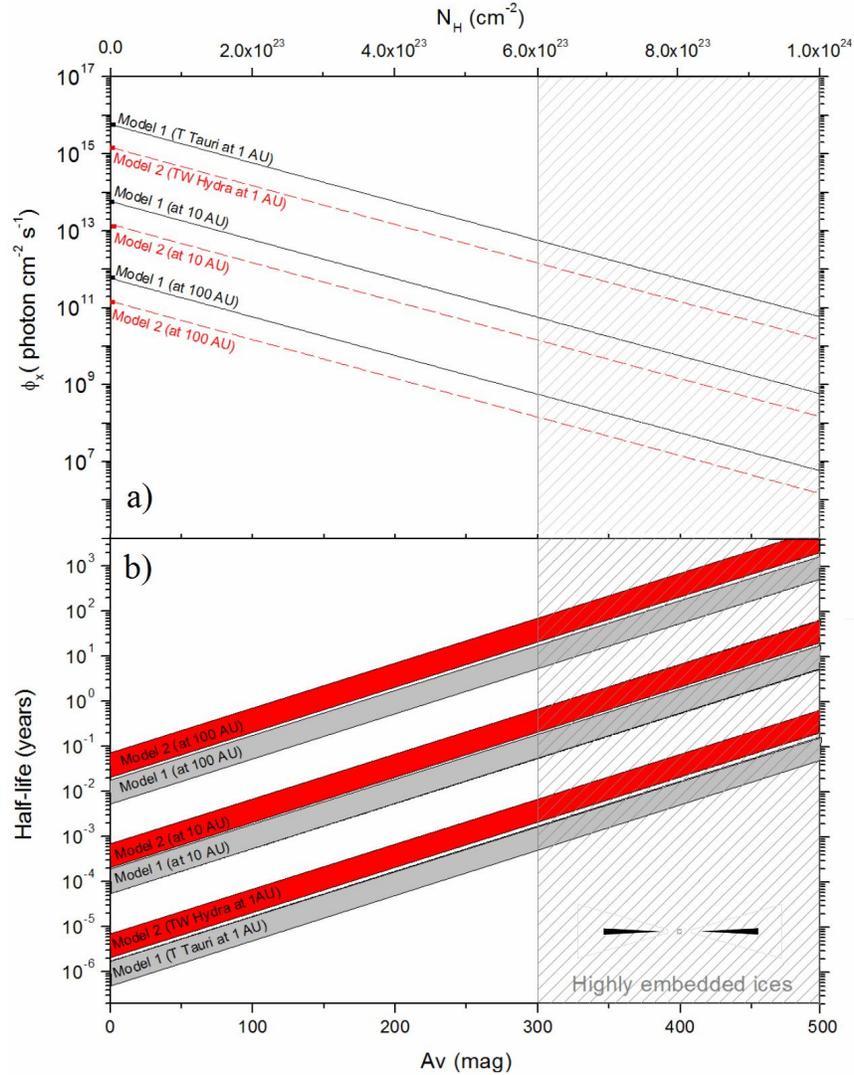

*Figure 8 - a) Estimative of photon flux (6-2000 eV) as function of visual extinction considering two models of soft X-ray ionization of YSOs: Model 1 (adapted from the model of Siebenmorgen & Krügel (2010) at 1 AU); Model 2 (adapted from Fantuzzi et al. 2011 at 1 AU). b) Range of half-lives of the studied ices (considering the destruction cross section in the range between 2 - 7 x 10$^{-18}$ cm$^{-2}$) as a function of visual extinction for both ionization models. See details in text.*

Considering the determined destruction cross section range (2-7 x 10$^{-18}$ cm$^{-2}$) in this work, given in Table 4, and the flux dependence with the visual extinction described above, we estimated the half-life of the studied ices in the vicinity of YSOs. The range of half-lives as function of visual extinction considering both X-ray ionization models of YSOs is seen in Figure 8b. Following the calculations, we observe that estimated half-lives derived in this work are stunningly low compared with the lifetime of YSOs of ~$10^5$-$10^6$ years (e.g. Dunham et al. 2014). However, if we considered that the denser regions of protostelar disk may have visual extinction of hundreds of magnitudes (highly embedded ices), such half-lives will increase considerably. Such regions with extremely high visual extinction may be found in massive YSOs, as discussed in the works of Stark et al. (2006), Boogert & Ehrenfreund (2004), and Ivezic & Elitzur (1997). Another important issue that also affects the amount and distribution of X-rays in the embedded ices is the location in which they are produced, and if their source has an extended profile or not (related with outflow and its jets).

Mendoza et al. (2013) have also employed dissociation cross section data in an attempt to evaluate the half-lives of frozen molecules illuminated by soft X-rays inside YSOs environments (e.g. TW Hydra). The authors have measured the destruction of frozen pyrimidine at 130 K by photons from 394 to 427 eV (low-energy soft X-rays). The obtained values for the half-lives at 100 AU (considering n (H) < $10^5$ cm$^{-3}$) are comparable with values determined in this study.

As observed from Table 4, the determined the destruction cross section values are within a factor of ≤ 2 from each other, indicating a small dependence on the ice temperature in this range, except for $NH_3$ at 50 K that is a factor of 3 larger. The results show that in both soft X-ray field models (TW Hydra or Typical T Tauri stars), the water and $SO_2$ molecules present in the warmer ices (90 K) are more sensitive to soft X-rays than ammonia and $CO_2$. However, for colder ices (50 K), the most sensitive species is $NH_3$. Additionally, the low value of half-lives estimated for both models reinforce that the idea of soft X-rays are indeed a very efficient source of molecular dissociation in such environments.

As mentioned previously, besides the direct heating promoted by ionizing field at ices in the circumstellar environment, they are also subjected to other heating processes such as grain-grain collision. Additionally, mass movements such as turbulence or convection inside the protostellar disc may induce a migration of grains from cold regions to hot regions (outwards). Therefore, along the evolution of YSOs, some ice grains embedded in the protostellar disk can be heated up to temperatures around 300 K (or higher), allowing only non-volatiles species to remain in the solid phases. An investigation about the influence of such extra heating (2 K min$^{-1}$ up to the room temperature) induced by non ionizing source in the irradiated ices was also performed, and the main results are discussed in the Appendix. During this final stage (thermal heating), the physicochemical changes of the sample were also continuous monitored by IR spectroscopy and the spectra are shown in Figure A2. This secondary stage of thermal heating delivered different species to gas phase, when compared to the first stage of heating (beginning of the experiments), because new species, produced by irradiation of the ices, are now available to react during molecular diffusion and desorption processes. Therefore, we expect that the gaseous environment around grains (molecular gas) inside the protostellar molecular core will be quite different when compared with the gas composition (envelope and inside disks) present in the late stages of protostellar environment.

Figure 9 presents the spectra of the organic residue produced for the studied ices initially containing the mixture $H_2O:CO_2:NH_3:SO_2$ (roughly 10:1:1:1) at two temperatures (50 and 90K) after irradiation and heating to room temperature. Although the ices have roughly similar initial composition when deposited at 13 K, the different temperature of the experiments allows chemical differentiation (enhanced during irradiation phase). The observed chemical differences between the studied ices (observed also at the residues at 300 K) can be justified by different reaction mechanisms, reaction rates, and desorption temperatures (and rates) of daughter species, all of them triggered by different temperatures.

For comparison purpose, the reference spectrum of two sulfur-containing amino acids D-Methionine (NIST webbook) and L-Cystine (Cataldo et al. 2011), the simplest proteinaceous amino acid glycine (Pilling et al. 2013), and the nucleobase adenine (Pilling et al. 2011a), all at room temperature, are also shown in Fig 8. Vertical lines indicate tentative matches between some peaks in the infrared spectra of the residues with peaks in the reference spectra. Asterisks indicate the peaks that only exist in residuals of E50K experiment. Such peaks may have indirect relation with volatile species (e.g. CO, $N_2$, $H_2$), produced due to the irradiation. These molecules may be trapped in the ice, thus participating in some chemical reactions to produce new species. The results indicate that some bands may have some contribution of peaks of these species. However, additional studies should be done employing other analytical chemical techniques, such as gas chromatography (GC/MS) or nuclear magnetic resonance (NMR), to precisely identify and quantify them among the chemical inventory of the residues. Therefore, the results suggest that complex molecules may be produced by soft X-rays in such extreme and cold environments.

Finally, besides the implications on the ice around YSOs, the current study may also be employed to draw the influence of soft X-rays as well as other energetic ionizing radiation in the physicochemistry of other $SO_2$-containing ices such as the surface of the Moon Europa (eg. Schriver-Mazzuoli 2003b; Carlson et al. 2009).

Pilling et al. (2009), in a similar study, also using VUV photons and soft X-rays of the Brazilian synchrotron light source, detected adenine employing ex-situ GC/MC and RMN analysis of the residue after the irradiation of an icy sample. The authors irradiated a mixture of $N_2$ and $CH_4$ in an attempt to simulate the effect of ionizing photons in the aerosol analogs of the upper atmosphere of Titan, a moon of Saturn.

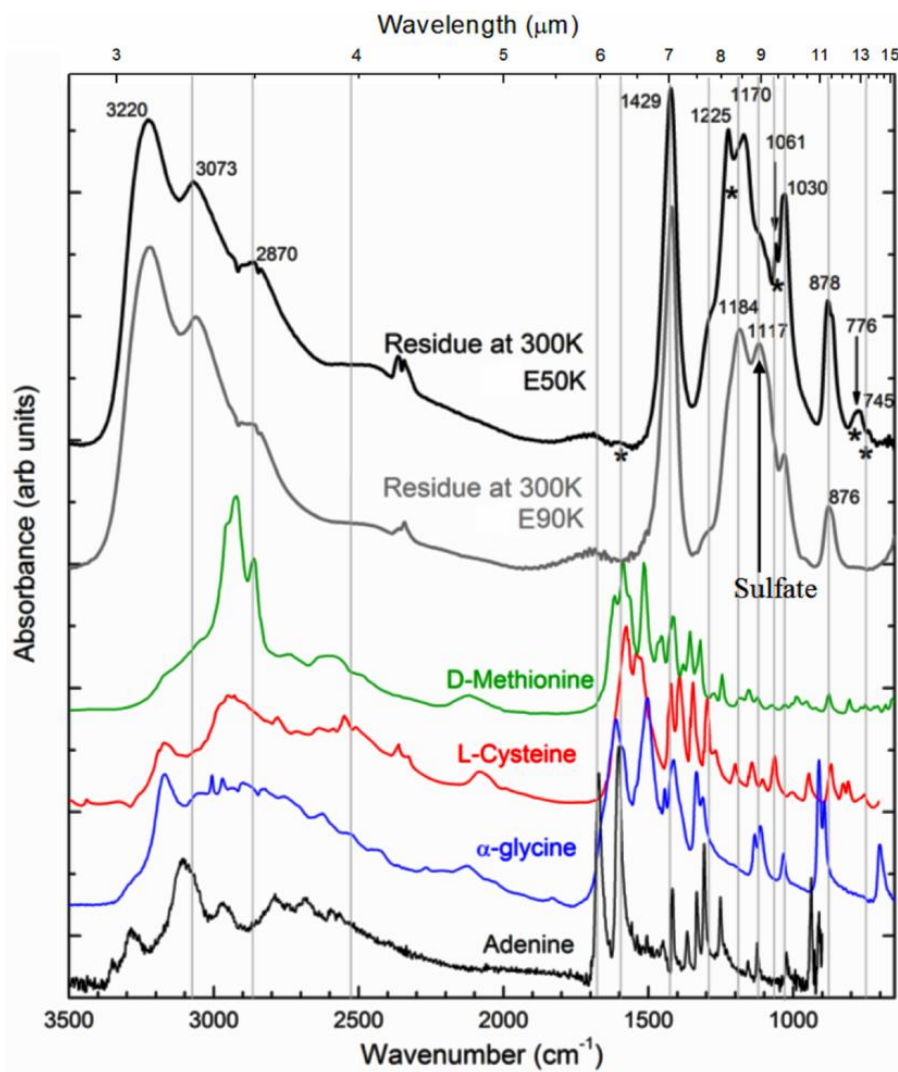

*Figure 9 - Comparison between the two residues of the studied ices containing the mixture $H_2O:CO_2:NH_3:SO_2$, after irradiation and heating to room temperature. Reference spectrum at room temperature of two sulfur-containing amino acids D-Methionine (NIST webbook) and L-Cystine (Cataldo et al. 2011), glycine (Pilling et al. 2013), and adenine (Pilling et al. 2011a) are also shown. Vertical lines indicate tentative matches between some peaks in the infrared spectra of the residues with peaks in the reference spectra of selected reference sample. Numbers indicate the apex of selected peaks. Asterisks indicate the peaks that only exist in the residue of the E50K experiment.*

## 5 CONCLUSIONS

In this work, we present for the first time the results of the experimental investigation of the effects produced by broadband soft X-rays (plus fast photoelectrons, low-energy induced secondary electrons, and a minor component of VUV photons) with energies from 6 to 2000 eV, on the frozen mixtures containing $H_2O:CO_2:NH_3:SO_2$ (10:1:1:1) at 50 K and 90 K. The experiments are an attempt to simulate the real photochemical processes induced by energetic photons in $SO_2$-containing ices present in cold environments inside surrounding young stellar objects (largely illuminated by soft X-rays). The measurements were performed using a high vacuum portable chamber from the Laboratório de Astroquímica e Astrobiologia (LASA/UNIVAP) coupled to the spherical grating monochromator (SGM) beamline at the Brazilian Synchrotron Light Source (LNLS) in Campinas, Brazil. *In-situ* sample analyses were performed by a Fourier transform infrared spectrometer. Our main results and conclusions are the following:

i) The dissociation cross section of parental species due to the irradiation of broadband soft X-rays were in the order of $2\text{-}7\times10^{-18}$ $cm^2$. For the $CO_2$ and $SO_2$, the temperature of the ices in the studied range did not significantly influence the dissociation cross sections. The determined values were around $3\times10^{-18}$ $cm^2$ for the $CO_2$ and $4\times10^{-18}$ $cm^2$ for the $SO_2$. The ammonia was the species most affected by the irradiation temperature. The destruction cross section at 50 K was roughly 3 times higher than the value obtained at 90 K. This may happen due to the opening of specific reaction pathways involving $NH_3$ at low temperatures. Curiously, an opposite behavior was observed for the other species that have the higher dissociation cross section at 90 K.

ii) The infrared spectra of irradiated samples presented the formation of several organic molecules, including nitriles, acids, $H_2SO_4$, $SO_3$, and other organic compounds. The formation of $OCN^-$ was much enhanced with the increasing temperature of the samples. When applied to the circumstellar environment of YSOs, this suggests that the abundance of $OCN^-$ maybe enhanced in the warmer regions (Av < 3), for example, in the external layers of the protoplanetary disk.

iii) The half-life of the studied parental species in the sample, extrapolated to YSO conditions, suggests that a chemical gradient between the cold and hot regions inside protoplanetary disk may also be triggered by incoming soft X-rays. The results show that for colder ice, the most sensitive species to X-rays was ammonia. For the warmer ice (T ~ 90 K), the most sensitive species was the water. The low value of half-lives obtained, employing two different models of radiation field of YSOs (typical T Tauri star and TW Hydra), reinforces that soft X-rays are indeed an very efficient source of molecular dissociation in such environments.

iv) During the heating to room temperature after the irradiation of both samples, the sulfur dioxide remained visible in the infrared spectrum (trapped in the ice) until temperatures of 195-200 K. Such high temperature, higher than the sublimation temperature of water ice, indicates that at least a fraction of $SO_2$ must remain in ices toward YSOs after an extra heating produced, for example, by an energy delivered during grain-grain collision or cosmic-ray heating. Due to this difference in the sublimation temperature between $H_2O$ and $SO_2$, in such possible heating scenarios, the molecular ratio $SO_2/H_2O$ may increase the production of sulfurous-bearing molecules, such as $SO_3$, and also increase the probability of the sulfur-containing biomolecules. The IR spectrum of the residues after heating to room temperature presented some bands that match peaks in the reference spectra of selected sulfur-bearing amino acids, glycine, and adenine. However, additional experiments should be done employing other analytical chemical techniques, such as gas chromatography (GC/MS) or nuclear magnetic resonance (NMR), to precisely identify and quantify them among the chemical inventory of the residues.

The current experimental investigation reveals the importance of soft X-rays and induced electrons in the chemical modification of simulated astrophysical ices. Moreover, this study confirms previous results, which showed that the organic chemistry on the ices around YSOs could be very complex, temperature dependent, and extremely rich in prebiotic compounds (triggered by incoming radiation field).

**ACKNOWLEDGMENTS**

The authors acknowledge the Brazilian agencies FAPESP (Projects JP 2009/18304-0 and DR 2012/17248-2) and CNPq (Research fellowship 304130/2012-5) as well as the FVE/UNIVAP for financial support. We thank MSc. Will R. M. Rocha for fruitful discussions as well as the staff at UNIVAP and LNLS for technical support.

**REFERENCES**

Andrade, D. P. P., Boechat-Roberty, H. M., Pilling S., da Silveira, E. F., & Rocco, M. L. M. 2009, SurSc, 603, 3301

Akkerman, A., Breskin, A., Chechik, R., & Gibrekhterman, A., 1993, Secondary Electron Emission from Alkali Halides Induced by X-Rays and Electrons. In Ionization of Solids by Heavy Particles, Ed. R. A. Baragiola, NATO ASI Series 306, pp 359-380, Springer US

Almeida, G. C., Pilling, S., Andrade, D. P. P., et al. 2014, JPCC, 118, 6193

Andrade, D. P. P., Boechat-Roberty, H. M., da Silveira, E. F., et al. 2008, JPCC, 112, 11954

Baragiola, R. A., Famá, M. A., Loeffler, M. J., Raut, U., & Shi, J. 2008, NIMPB, 266, 3057

Bennett, C. J., & Kaiser, R. I. 2007, ApJ, 661, 899

Bernstein, M. P., Ashbourn, S. F. M., Sandford, S. A., & Allamandola, L. J. 2004, ApJ, 601, 365

Bergantini, A., Pilling, S., Nair, B. G., Mason, N. J., & Fraser, H. J. 2014, A&A 570, A120

Boechat-Roberty, H. M, Pilling, S. & Santos, A. C. F., 2005, A&A, 438, 915

Boechat-Roberty, H. M., Neves, R., Pilling, S., Lago, A. L., & de Souza, G. G. B. 2009, MNRAS, 394, 810

Boogert, A. C. A., Schutte, W. A., Tielens, A. G. G. M., et al. 1996, A&A, 315, L377

Boogert, A. C. A., Schutte, W. A., Helmich, F. P., Tielens, A. G. G. M., Wooden, D. H. 1997, A&A, 317, 929

Boogert, A. C. A., Hogerheijde, M. R., Ceccarelli, C., et al. 2002, ApJ, 570, 708

Boorgert A.C.A and Ehrenfreund P., 2004 ASPC 309, Astrophysics of dust, pp 547-573. Editors A. N. Witt, G. C. Claytion & B. T. Draine.

Boogert, A. C. A., Pontoppidan, K., Knez, C., et al., 2008, ApJ, 678, 985

Boogert A.C.A., Gerakines P.A., & Whittet D.C. B., 2015, ARAA, 53, in press (http://arxiv.org/pdf/1501.05317v2.pdf)

Cataldo, F., Ragni, P., Iglesias-Groth, S., & Manchado, A., 2011, JRNC, 287, 573

Carlson, R.W., Calvin, W., Dalton, J., et al. 2009 Europa's surface composition. In Europa, Eds. R.T. Pappalardo, W.B. McKinnon, and K.K. Khurana, pp 283–327, University of Arizona Press, Tucson, AZ

Casanova, S., Montmerle, T., Feigelson, E. D. & Andre P., 1995, ApJ, 439, 752

Caux, E., Kahane, C., Castets A., et al. 2011, A&A, 532, A23

Cazaux J., 2006, NIMPB, 244, 307

Cecchi-Pestellini, C., Ciaravella, A., Micela, G., & Penz, T. 2009, A&A, 496, 863

Collings, M. P., Anderson, M. A., Chen, R., et al. 2004, MNRAS, 354, 1133

Comito, C., Schilke, P., Phillips, T.G., et al. 2005, ApJS, 156, 127

Chen, Y.-J., Ciaravela, A., Munoz Caro G., et al. 2013, ApJ, 778, 162

Ciaravela, A., Munoz Caro, G., Jimenez Escobar A., et al. 2010, ApJL, 722, L45

Castro, A. R. B., Fraguas, G. B., Pacheco, J. G., et al., 1997, Commissioning of the SGM beamline, LNLS Tech Memorandum CT 06/97 (Memorando Tecnico), LNLS, Campinas, Brazil.

d'Hendecourt, L., & Allamandola, L. J. 1986, A&AS, 64, 453

Dunham M.M., Stutz A.M., Allen L. E., et al. 2014 The evolution of protostars: Insights from ten years of infrared surveys with Spitzer and Herchel. In Protostars and Planets VI. Eds. H. Beuther et al., pp 129-218. Univ. of Arizona, Tucson, AZ

Eagleson, M., 1994, In Concise Encyclopedia Chemistry, Walter de Gruyter & CO, New york.

Elsner, R. F., Gladstone, G. R. & Waite J. H., 2002, ApJ, 572, 1077

Fantuzzi F., Piling. S., Santos, A. C. F., et al. 2011, MNRAS, 417, 2631

d'Hendecourt, L. B., & Allamandola, L. J. 1986, A&ASS, 64, 453
Garozzo, M., Fulvio, D., Gomis, O., Palumbo, M. E., & Strazzulla, G. 2008, P&SS, 56, 1300
Gerakines, P. A., Schutte, W. A., Greenberg, J. M., et al. 1995, A&A, 296, 810
Gerakines, P. A., Whittet, D. C. B., Ehrenfreund, P, et al., 1999, ApJ 522, 357
Gerakines, P. A., Moore, M. H., Hudson, R. L. 2001, JGRE, 106, 33381
Gibb, E. L., Whittet, D. C. B., Schutte, W. A., et al., 2000, ApJ, 536, 347
Gibb, E. L., Whittet, D. C. B., Boggert, A. C. A., & Tielens, A. G. G. M. 2004, ApJS, 151, 35.
Goicoechea, J. R., Rodriguez-Fernandez, N. J., & Cernicharo, J. 2004, ApJ, 600, 214
Gueymard, C. A. 2004, SoEn, 76, 423
Gullikson, E. M., & Henke, B. L. 1989, PhRvB, 39, 1
Güver, T., & Özel, F. MNRAS, 2009, 400, 2050
Hendrix, A. R., Cassidy, T. A., Johnson, R. E., Paranicas, C., & Carlson, R.W. 2011, Icar 212, 736
Henke, B. L., Liesegang J., & Smith, S. D. 1979, PhRvB., 19, 3004
Henning, Th., & Semenov, D. 2013, ChRv, 113, 9016
Hüfner, S., 1995, In Photoelectron spectroscopy: principles and applications. Springer Verlag
Imanishi, K., Tsujimoto, M & Koyama, K., 2002, ApJ, 572, 300
Imanishi, K., Koyama K., & Tsuboi Y., 2001, ApJ, 557, 747
Ivezic Z., & Elitzur M., 1997, MNRAS 287, 799
Ivlev A.V., Röcker, T.B., Vasyunin, A., & Caselli P. 2015, ApJ, 805, 59
Kaiser, R. I. 2002, ChRv, 102, 1309
Keane, J. V., Tielens, A. G. G. M., Boogert, A. C. A., Schutte, W. A., & Whittet, D. C. B. 2001, A&A, 376, 254
Kerkhof, O., Schutte, W.A., & Ehrenfreund, P., 1999, A&A, 346, 990
Koyama, K., Hamaguchi, K., Ueno, S., Kobayashi, N., & Feigelson, E. D., 1996, PASJ, 48, L87
Khanna, R. K., Zhao, G., Ospina, M. J., Pearl, J. C., 1988, AcSpA, 44A, 581
Jacox, M. E., & Milligan, D. E., 1971, JChPh. 54, 919
Jimenez-Escobar A., Munoz Caro G. M., Ciaravella A., et al. 2012, ApJL, 751, L40
Johnson, R. E., & Quickenden T. I. 1997, JGR, 102, 10985
Lattanzi, V., Gottlieb, C. A., Thaddeus, P., et al. 2011, A&A 533, L11
Loeffler, M. J., & Hudson, R. L., 2013, Icar, 224, 257
Loeffler, M. J., & Hudson, R. L., 2010, GeoRL, 37, 19201
Loerting T., & Liedl, K. R., 2000, PNAS, 97, 8874
Maloney, P. R., Hollenbach, D. J., & Tielens, A. G. G. M., 1996, ApJ, 466, 561
Mendoza E., Almeida G. C., Andrade D.P.P, et al. 2013, MNRAS, 433, 3440
Moore, M. H. 1984, Icar, 59, 114
Moore, M. H., Hudson, R. L., & Carlson R.W. 2007, Icar, 189, 409
Öberg, K., Boogert, A.C.A., Pontoppidan, K.M., et al. 2011, ApJ, 740, 109
Opal, C. B., Beaty E. C., & Peterson W. K., 1972, ADNDT, 4, 209
Orlando, T. M., & Kimmel, G. A. 1997, SurSc, 390, 79
Palumbo, M. E. 2006, A&A, 453, 903
Patterson, G. W., Paranicas, C., & Prockter, L.M. 2012, Icar, 220, 286
Pilling, S., Santos, A. C. F., Boechat-Roberty, H. M. 2006, A&A, 449, 1289
Pilling, S., Neves, R., Santos, A. C. F., & Boechat-Roberty, H. M. 2007a, A&A, 464, 393
Pilling S., Andrade, D. P. P., Neves, R., et al. 2007b, MNRAS, 375, 1488
Pilling, S., Boechat-Roberty, H. M., Santos, A. C. F., de Souza, G. G. B., & Naves de Brito, A., 2007c, JESRP,156-158, 139
Pilling, S. et al., 2008 in Proceedings of the International Astronomical Union (Symposium 251), 4, 371. DOI:10.1017/S1743921308021996. Cambridge press.
Pilling, S., Andrade, D. P. P., Neto, A. C., Rittner R., & Naves de Brito, A. 2009, JPCA, 113, 11161
Pilling, S., Seperuelo Duarte E., da Silveira E. F., et al. 2010a, A&A, 509, A87
Pilling, S., Seperuelo Duarte E., Domaracka A., et al., 2010b, A&A, 523, A77

Pilling, S., Andrade, D. P. P., do Nascimento E. M., et al. 2011a, MNRAS, 411, 2214.
Pilling, S., Baptista, L., Andrade, D. P. P., & Boechat-Roberty, H. M., 2011b, AsBio, 11, 883
Pilling, S., Seperuelo Duarte E., Domaracka A., et al. 2011c, PCCP, 13, 15755
Pilling, S., & Andrade, D. P. P. 2012, In Employing Soft X-Rays in Experimental Astrochemistry, X-Ray Spectroscopy, Shatendra K. Sharma (Ed.), p. 185-218, ISBN: 978-953-307-967-7, InTech, DOI: 10.5772/29591
Pilling, S., Andrade, D. P. P., da Silveira, E. F., et al. 2012. MNRAS, 423, 2209
Pilling, S., Mendes L. A., Bordalo V., et al. 2013 AsBio, 13, 79
Pilling, S., Nair B. G., Escobar A., Fraser, H., Mason, N. 2014, EPJD, 68, 58
Ramaker, D. E., Madey, T. E., & Kurtz , R. L. 1988, PhRvB, 38, 2099
Rocha, W. M., & Pilling, S. 2015, ApJ, 803, 18
Rodrigues, A. R. D., Craievich, A. F., & Goncálves da Silva, C. E. T. 1998, J. Synchrotron Rad., 5, 1157.
Schilke, P., Benford, D. J., Hunter, T. R., Lis, D. C., & Phillips, T. G. 2001, ApJS, 132, 281
Schleicher, D. R. G., Spaans M., & Klessen, R. S., A&A, 2010, 513, A7
Schriver-Mazzuoli, L., Chaabounia, H., & Schriver, A. 2003a, JMoSt, 644, 151
Schriver-Mazzuoli, L., Schriver, A., & Chaabouni, H. 2003b, CaJPh, 81, 301
Shimanouchi, T., 1972, Nat. Stand. Ref. Data Ser, Nat. Bur. Stand. NSRDS-NBS 39, USA.
Shen, C. J., Greenberg, J. M., Schutte, W. A., & van Dishoeck, E. F. 2004, A&A, 415, 203
Siebenmorgen, R., & Krügel, E., 2010, A&A, 511, A6
Snyder, L. H., Hollis J. M., Ulich B. L., et al. 1975, ApJ, 198, L81
Stark, D.P., Whitney, B.A., Stassun, K., & Wood, K. 2006, ApJ, 649, 900
Strazzulla, G. 2011, NIMPB, 269, 842
Tielens A. G. G. M., & Hollenbach D. 1985, ApJ, 291, 722
Turner, B. E. 1994, ApJ, 430, 727
Zasowski, G., Kemper, F., Watson, D. M., et al. 2009, ApJ, 694, 459
Zhang R., Wooldridge P. J, Abbatt J. P. D., & Molina, M. J., 1993, JPhCh, 97, 7351
Zheng, W., Jewitt, D., & Kaiser, R. I. 2009, ApJS, 181, 53

**APPENDIX A- THERMAL HEATING BEFORE AND AFTER IRRADIATION.**

**A.1   HEATING PHASE BEFORE IRRADIATION: SAMPLE PREPARATION**

Figure A1 presents different spectra of the experiment labeled E50K and E90K before irradiation phase, during the slow heating (2 K min$^{-1}$) from 13 K (temperature of the frozen sample production) up to the temperature of irradiation phase: 50 K and 90 K, respectively. Figure A1 bottom panels show an expanded view from 1800 to 610 cm$^{-1}$. Labels identify specific infrared bands of the molecular species ($H_2O$, $CO_2$, $NH_3$, $SO_2$) in the spectra. Such bands are associated with different molecular vibration modes of the frozen species (e.g. Pilling et al. 2010a, 2010b, 2011a). In all panels, the uppermost spectrum is the warmer spectrum. As observed in the figure for the E50K experiment, apparently, the heating from 13 to 50 K did not affect the infrared spectrum of the sample. However, for the E90K experiment, some chemical reactions were triggered during the heating process (mainly when temperatures reached around 80 K). The vertical bars in these figures indicate the new infrared bands associated with the produced species in the ice sample due to the heating. The new infrared peaks identified in the spectrum appeared at the wavenumbers 1552, 1493, 1395, 1295, 1035, 1011, 1011, 957, and 826 cm$^{-1}$.

Following Schriver-Mazzuoli et al. (2003b), the peak at 1395 cm$^{-1}$ can be attributed to sulfur trioxide ($SO_3$). The peak around 1552cm$^{-1}$ is tentatively attributed to carboxylate ion (RCOO$^-$), or even amino acid zwitterions (Shimanouchi, 1972). The feature at 1035 cm$^{-1}$ was attributed to ozone ($O_3$) and/or to hydrogen sulfite ion (bissultite ion) $HSO_3^-$ by Moore et al. (2007) in similar experiments. The presence of sulfuric acid ($H_2SO_4$) and the disulfite ion ($S_2O_5^{-2}$) can be attributed to the small peak at 957 cm$^{-1}$ (e.g.

Moore et al. 2007, Loeffler & Hudson 2013). The disulfite ion is a dimer of the bisulfite ion ($HSO_3^-$) and may also be transformed back in the bisulfite ion in the presence of acids of free protons (Eagleson, 1994). The disulfite ion also arises from the addition of sulfur dioxide to the sulfite ion ($SO_3^{-2}$). A similar chemical alteration was also observed by Loeffler & Huddson (2010) during thermal processing of SO2-containing ices.

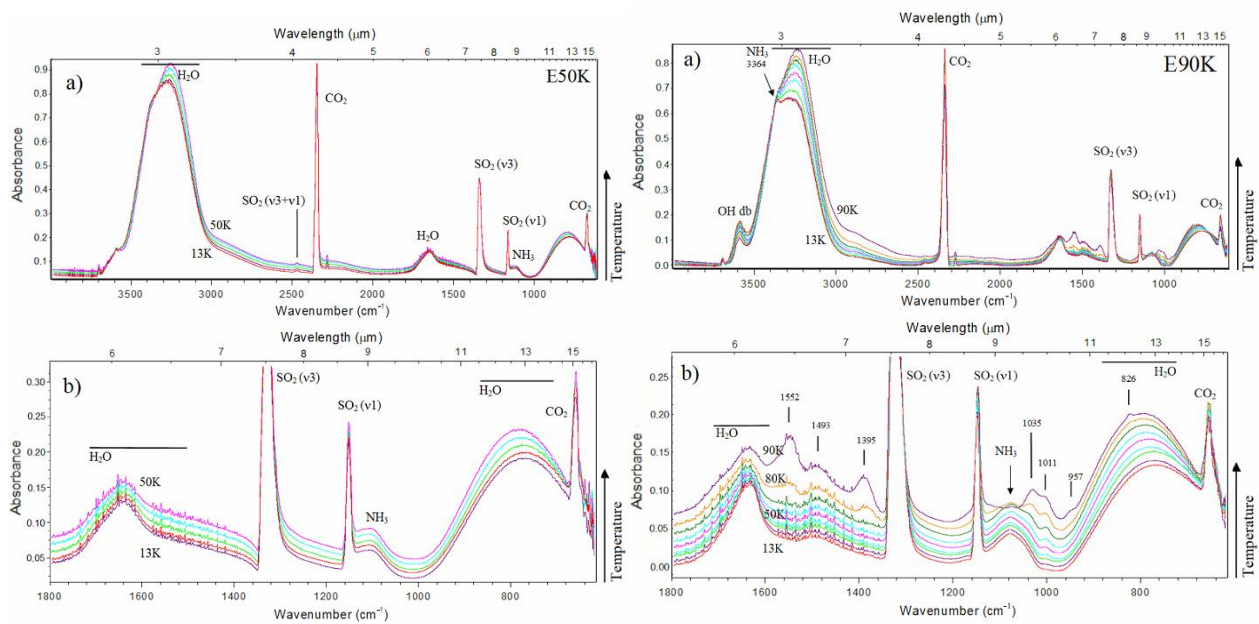

*Figure A1- a) Evolution of the unirradiated infrared spectrum sample of the E50K (left) and E90K (right) experiment during the thermal heating (2 K min$^{-1}$) from the initial temperature of 13 K to the temperature of irradiation phase. b) Expanded view from 1800 to 610 cm$^{-1}$.*

It is import to note that, the water ratio employed in the studied ices is small compared with the one present in astrophysical ices, and the observed chemical changes associated with sulfur dioxide reaction, triggered by temperature enhancement, may have a minor contribution in real astrophysical scenario. Future experiments containing different abundance of water in the ice may help to clarify this issue. During the initial heating of the unirradiated samples, the chemical ratio of the parental species changed slightly in comparison with the samples at 13 K. This occurs due to the sublimation of the weakly bonded molecules, and became more evident in the E90K experiment, which also allowed phase transition of the trapped ammonia (Zheng et al. 2009) and $SO_2$ (Schriver-Mazzuoli et al. 2003a). In addition, the presence of strong OH dangling bonds (OHdb) at around 3610 cm$^{-1}$ indicates that E90K sample presented a higher degree of porosity (see also Palumbo 2006; Pilling et al. 2010a, 2010b). No significant change in this band was observed during the heating from 13 to 90 K.

## A.2 HEATING PHASE AFTER IRRADIATION: RESIDUE PREPARATION

Figure A2a presents the IR spectra collected at specific temperatures during the heating of the irradiated sample at 50 K (E50K experiment) at the final fluence of 3.9x10$^{18}$ photons cm$^{-2}$, obtained after 648 min of continuous irradiation with synchrotron light. During this heating stage, some infrared bands vanish as a result of molecular sublimation and chemical changing. However, other IR bands appear as well, illustrating the formation of new species. For example, the following bands appeared at: 1420 cm$^{-1}$ at the temperature of 160 K; 1250 cm$^{-1}$ at 172 K; 880 cm$^{-1}$ and 780 cm$^{-1}$ at 183 K; 1040 and 1230 cm$^{-1}$ at 265 K. Such bands and their tentative molecular assignments are listed in Table 3. The second uppermost spectrum in this figure represents the residue at room temperature, and the uppermost spectrum represents the residue after a second cooling-down cycle to 13 K. In this re-cooled sample spectrum, some bands observed in the residue at 300 K have changed to a sharper profile, indicating that some phase transitions may have occurred. Figure A2b presents IR spectra collected at specific temperatures during the heating of the bombarded simulated sample at 90 K in the final fluence of 3.6 x10$^{18}$ photons cm$^{-2}$, obtained after 600 minutes of continuous irradiation with synchrotron light. During the heating, $SO_2$ and $CO_2$, which are the remaining parental species, desorbed completely in temperatures above 200 K and 260 K, respectively. In addition, at temperatures above 200 K, new features arose in the IR spectrum in the wavenumbers 880, 1429, 2871, and 3073 cm$^{-1}$. A list of the new bands observed in the spectra during the heating and their tentative molecular assignments are shown in Table 3. As discussed before, $SO_3$ was one of the molecules produced during the irradiation. This species was also observed in the residue at room temperature (e.g. 828, 1060, 1215, 1396 cm$^{-1}$). In a similar set of experiments involving the processing of $SO_2$-rich ices by fast protons, the $SO_3$ molecule was also observed (Moore 1984; Strazzulla 2011).

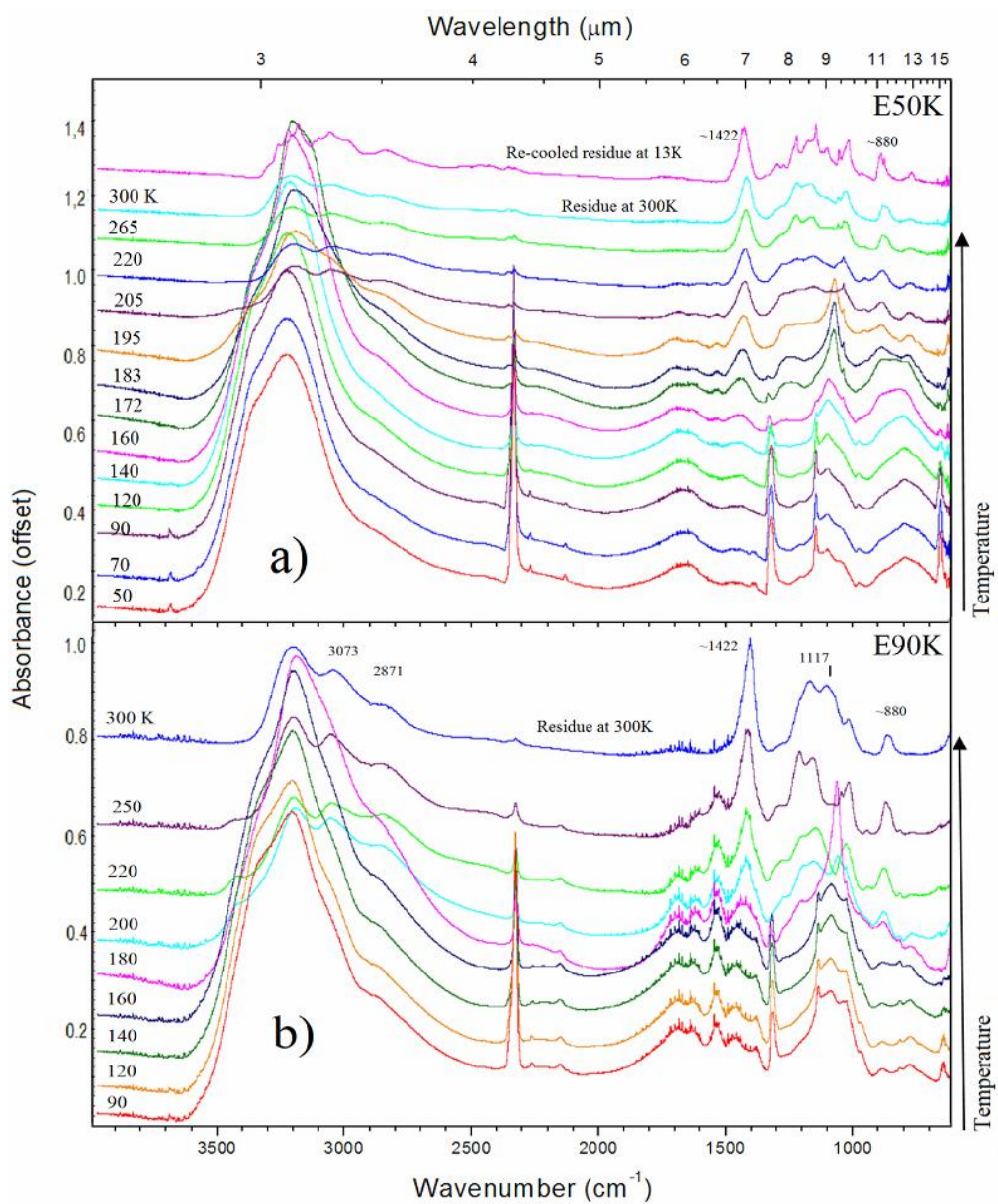

*Figure A2 - Changes in the IR spectrum of the irradiated samples during the slow heating up to the room temperature. a) Irradiated sample at 50 K (final fluence of 3.9 x10$^{18}$ photons cm$^{-2}$) during heating to 300 K. The uppermost spectrum is the residue at 300 K after cooling down again to 13 K (re-cooled sample). b) Irradiated sample at 90K (final fluence of 3.6x10$^{18}$ photons cm$^{-2}$) during the heating to 300 K. See details in the text.*

Because water is the most abundant species in the current simulation, during the heating, minor parental species are still trapped in the matrix, and besides some desorption at specific temperatures, most of them also desorbed at the desorption temperatures of water molecules, around 160-190 K (see discussion also in Collings et al. 2004). Following the authors, the desorption temperatures of pure $CO_2$, $NH_3$, and $SO_2$ ices at low pressure are around 80, 90, and 100 K, respectively. Therefore, during the heating stage, a fraction of such species is expected to diffuse, react, and also desorbed from the sample at these specific temperatures.

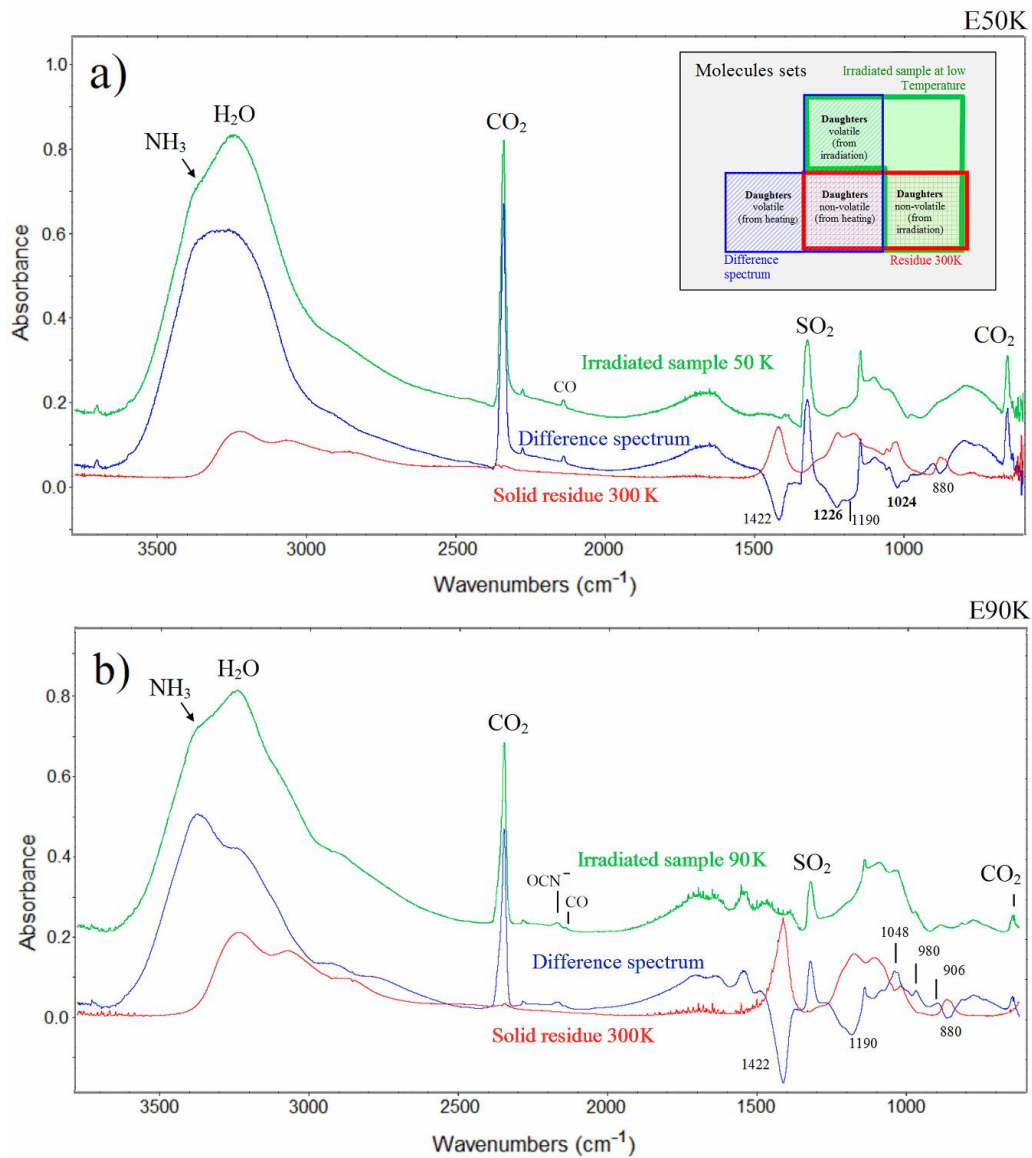

*Figure A3 - Comparison between infrared spectrum of solid residue at 300 K (non-volatile molecules) and the spectrum of the irradiated sample at low temperature. Difference spectrum (obtained from the subtraction between the spectrum of the irradiated sample at low temperature and the IR spectrum of the residue at 300 K). a) Experiment E50K). b) Experiment E90K). Inset figure indicates the family of species observed in each spectrum. See details in the text.*

The infrared spectra of irradiated samples during the final heating stage to room temperature helps us to understand what may happen in the surface chemistry when these ices are exposed to external or internal heating process. When associated with Europa ices, such heating mechanisms may be, for example, ruled by tidal forces and/or asteroid/comet impact. As discussed before, for the ices in the disk of YSOs, this heating may be produced by grain-grain collision, delivered comic ray energy, or by the absorption of IR and visible photons from the host star. In both situations, if the sample temperature is heated to values higher than 180-200 K, the chemistry of this region will change drastically (mainly due to the water sublimation and the opening of reaction pathways on the ice). As observed in Figure A2b, this seems to be more enhanced for the E90K experiment than for the 50 K experiment, when we make a comparison between spectra below 180 K with that obtained above 200 K.

Figures A3a and A3b present a comparison between the IR spectra of the solid residue at 300 K (non-volatile molecules) and the IR spectra of the irradiated sample at low temperature, for the experiment E50K and E90K, respectively. The difference spectrum (obtained from the subtraction between spectrum of the irradiated sample at low temperature and the spectrum of the residue at 300 K) is also shown. This difference spectrum shows bands with positive and negative area. Such bands represent different sets of molecular species: positive profiles indicate volatiles species (parental species and daughter species produced due to the irradiation and also during the heating phase); Negative profiles indicate only the vibration modes of the non-volatiles species produced during the heating phase (e.g. bands at 1422, 1190, and around 880 cm$^{-1}$).

The inset in Figure A3a indicates the family of species observed in each spectrum. For example, in spectra of the irradiated samples at low temperature, we observe infrared features of the volatile parental species combined with features from volatiles and non-volatiles daughter species produced. In the infrared spectrum of the solid residue at 300 K we observe the IR features of the non-volatile species, produced due to the irradiation, and also the non-volatiles species produced during the slow heating to room temperature after the irradiation phase.

The results indicate that during the final (thermal) heating, we observe only in the E50K experiment non-volatile species with peaks in the infrared spectrum at 1226 and 1024 cm$^{-1}$. This suggests that besides the particularities in the surface chemistry induced by irradiation, there is an additional factor that is clearly induced by thermal and non-ionizing heating.